%% file: aanda.tex
\documentclass{aa}  

\usepackage{graphicx}
\usepackage{txfonts}
\usepackage{hyperref}
\usepackage{orcidlink}
\definecolor{mylightblue}{rgb}{0.19,0.55,0.91}
\hypersetup{colorlinks,breaklinks,
            urlcolor=mylightblue,citecolor=violet,
            linkcolor=mylightblue}
\graphicspath{{figures/}{../figures/}}
\usepackage{graphicx}

\input{include_tex/pacchetti}

\input{include_tex/journals}

\graphicspath{{figures/}{../figures/}}
\usepackage{newtxtext,newtxmath}
\usepackage[T1]{fontenc}
\usepackage{graphicx}	% Including figure files
\usepackage{amsmath}	% Advanced maths commands
\usepackage{amssymb}	% Extra maths symbols
\usepackage{hyperref} % link in bibliografia
\usepackage{natbib}
\usepackage{cleveref} % pacchetto per le citazion
\newcommand{\msun}{\ensuremath{\: \rm{M_{\odot}}}}
\newcommand{\erg}{\ensuremath{\: \rm{erg}}}

\def\CRIMSONS{\texttt{CRIMSONS} }
\newcommand\addition[1]{#1}

\usepackage{txfonts}
\begin{document}

   \title{CRIMSONS: An Online Tool for Modeling Chemical Enrichment with Stochastic IMF Sampling}
    % title for short
   \titlerunning{CRIMSONS}

   \author{ 
          M. Rossi\orcidlink{0000-0001-6887-2663}\inst{1,2} \fnmsep\thanks{\href{mailto:martinarossi2707@gmail.com}{martinarossi2707@gmail.com}}
          \and
          L. Querci\orcidlink{0009-0006-7675-2614}\inst{3,4}
          \and
          S. Ciabattini\orcidlink{0009-0000-3803-786X}\inst{3,4}
          \and
          S. Salvadori\orcidlink{0000-0001-7298-2478}\inst{3,4}
          \and
          I. Vanni\orcidlink{0000-0001-9647-0493}\inst{3}
          \and
          D. Massari\orcidlink{0000-0001-8892-4301}\inst{2}
          \and
          E. Ceccarelli\orcidlink{0009-0007-3793-9766}\inst{5}
          \and
          D. Romano\orcidlink{0000-0002-0845-6171}\inst{2}
          \and 
          V. Gelli\orcidlink{0000-0001-5487-0392}\inst{6}
          \and
          R. Pascale\orcidlink{0000-0002-6389-6268}\inst{2}
          \and
          A. Mori\orcidlink{0009-0003-0816-2880}\inst{3,4}
          \and
          E. Rusta\orcidlink{0009-0006-4326-6097}\inst{3,4}
          \and
          I.Koutsouridou\orcidlink{0000-0002-3524-7172}\inst{3}
          \and 
          {\'A}. Sk{\'u}lad{\'o}ttir\orcidlink{0000-0001-9155-9018}\inst{3}
          \and
          L. Magrini \orcidlink{0000-0003-4486-6802}\inst{4}
          \and
          R. E. Giribaldi \orcidlink{0000-0002-9420-560X}\inst{4}
          \and
          J. Schiappacasse-Ulloa \orcidlink{0000-0002-2179-9363}\inst{4}}

  \institute{
    Dipartimento di Fisica e Astronomia, Alma Mater Studiorum, Università di Bologna, Via Gobetti 93/2, 40129 Bologna, Italy
    \and
    INAF, Osservatorio di Astrofisica e Scienza dello Spazio, Via Gobetti 93/3, 40129 Bologna, Italy
    \and
    Dipartimento di Fisica e Astrofisica
    Univerisità degli Studi di Firenze,
    via G. Sansone 1, 
    Sesto Fiorentino, Italy
    \and
    INAF, Osservatorio Astrofisico di Arcetri, Largo E.Fermi 5, 50125, Firenze, Italy
    \and
    Kapteyn Astronomical Institute, University of Groningen, Landleven 12, 9747 AD Groningen, the Netherlands 
    \and
    Cosmic Dawn Center (DAWN), Denmark
    \and
    Niels Bohr Institute, University of Copenhagen, Jagtvej 128, 2200 København N, Denmark
    }

   \date{Received month XX, 2025; accepted Month XX, 2025}

  \abstract{
{Chemical enrichment is a universal process shaped by several interconnected factors, including stellar feedback, the nature of chemical polluters, and the underlying distribution of stellar masses. Three key ingredients are often difficult to treat simultaneously in chemical-evolution models: the stochastic sampling of the stellar Initial Mass Function (IMF), uncertainties in stellar nucleosynthesis yields, and pre-enrichment from primordial Population III (Pop III) stars. In low star-formation rate environments ($\mathrm{SFR} \lesssim 10^{-3},M_\odot,\mathrm{yr}^{-1}$), incomplete IMF sampling makes enrichment sensitive to the random presence or absence of massive supernova progenitors. Here, we present \CRIMSONS, the first online tool specifically designed to model chemical enrichment while self-consistently incorporating these three ingredients. The framework allows users to explore different IMFs, stellar population masses and metallicities, and multiple stellar yield sets. It follows the evolution of individual stars and tracks 30 elements (H--Zn), including contributions from supernovae, Asymptotic Giant Branch stars, and Type Ia supernovae. \addition{The current implementation adopts a closed-box framework, neglecting gas flows and spatial mixing, thus providing a controlled setup to isolate the effects of IMF sampling, stellar yields, and Pop III pre-enrichment}. With \CRIMSONS, we show that incomplete IMF sampling produces large ($\geq1$ dex) abundance scatter, particularly in low-mass stellar populations. Different yield prescriptions introduce element-dependent variations, while some abundance ratios remain comparatively robust across models. Finally, comparison between models including primordial enrichment and those adopting only a metallicity floor shows that Pop III pre-enrichment can significantly affect subsequent chemical evolution. \CRIMSONS\ therefore provides a controlled framework for quantifying the relative impact of stochastic IMF sampling, nucleosynthetic uncertainties, and primordial pre-enrichment on predicted abundance patterns.
}
}

\keywords{ Population III stars --- Population II stars  --- Chemical abundances --- \href{url:martina-rossi.it/crimsons.html}{martina-rossi.it/crimsons.html}}
   \maketitle
%
%-------------------------------------------------------------------
%     INTRODUCTION
%-------------------------------------------------------------------

\section{Introduction}
Chemical enrichment is a fundamental and universal process that shapes the evolution of astrophysical systems across cosmic time, from the formation of the first galaxies to the present-day Universe. The chemical composition of stars and gas retains a fossil record of stellar feedback, nucleosynthesis, and star formation, encoding key information on the physical processes governing galaxy evolution (e.g. \citealt{tinsley79, freeman02} ). Extracting this information from observed abundance patterns requires cosmological chemical evolution models that self-consistently connect stellar populations to the coupled effects of star formation, stellar feedback, and nucleosynthetic processes. Theoretical models which include the chemical enrichment commonly relies on simplifying assumptions. In particular, three key physical aspects are often treated separately or highly approximated: (i) the (pre-)enrichment from primordial metal-free Population~III (Pop~III) stars; (ii) the intrinsic discrete nature of the stellar Initial Mass Function (IMF) of stars; and (iii) the complex nature of stellar nucleosynthesis and end-of-life mechanisms, which translate into large theoretical uncertainties associated with stellar yields. Oversimplifying, or neglecting, the treatment of any of these aspects might heavily hampers our ability to reproduce the observed chemical peculiarity and diversity, especially at the low metallicity.

The first physical aspect that is often approximated is the role played by Pop~III stars in shaping the early chemical enrichment of galaxies. Pop~III stars, which are believed to have formed in metal-free minihaloes at $z \sim 20$–$30$ \citep[e.g.][]{abel02, bromm13}, were responsible for injecting the first heavy elements into the interstellar medium (ISM), setting the initial chemical conditions for subsequent star formation \citep[e.g.][]{klessen23}. Although their physical properties, such as the IMF, the supernovae (SNe) explosion energies, and the dominant SNe channels, remain highly uncertain, their imprint most likely shaped the early chemical enrichment of galaxies \citep[e.g.][]{salvadori07, hartwig15, magg19, rossi21, koutsouridou23, pagnini23}. The nucleosynthetic signatures of these primordial sources would be preserved in the atmosphere of long-lived low-mass Population II/I (Pop~II/I) stars that formed from a gas polluted by one or a few Pop III SNe, making very metal-poor stars powerful fossil records of the primordial chemical enrichment \citep[e.g.][]{beers05, iwamoto05, ishigaki14, frebel15, ji16,frebel18, ji20, skuladottir15, ezzeddine19, skuladottir21, placco21, vanni23, rossi23, bonifacio25}. However, many theoretical models and numerical simulations adopt a fixed metallicity floor (typically at $Z_{\rm ISM} \sim 10^{-3} Z_\odot$) above which Pop~II star formation is triggered \citep[e.g.][]{tornatore07,maio10,wise12, jaacks18}, thus underestimating the impact of the pre-enrichment from Pop~III stars. Such an approach may does not allow to reproduce the element-by-element abundance patterns observed in metal-poor stars and systems, and prevents from a realistic description of the chemical initial conditions inherited by later stellar generations.

A second major limitation is the assumption of a fully sampled IMF, which is justified when the star formation rate (SFR) is sufficiently high (SFR $\gtrsim 10^{-2},M_\odot\mathrm{yr}^{-1}$). On the contrary, in low-mass and poorly star-forming environments, such as ultra-faint dwarf galaxies, metal-poor star clusters, and early star-forming minihaloes, the theoretical IMF would not be statistically populated \citep{kroupa03, carigi08, leaman12, weidner13, debennessuti17, pagnini23}. As a consequence, the chemical enrichment becomes inherently stochastic as the presence or absence of even a single massive star might dominate the enrichment history, producing large variations in the elemental abundance ratios. In such cases, the chemical enrichment must be described as a discrete and probabilistic process, where individual stellar events, rather than IMF-averaged yields, govern the composition of the ISM. This stochasticity is widely recognized as a key driver of the observed scatter of chemical properties of metal-poor stellar populations \citep{vanni23,rossi24_1, rizzuti25}, yet only a limited number of modeling frameworks allow for a systematic exploration of IMF sampling effects \citep[e.g.,][]{da_silva_slugstochastically_2012, krumholz_slug_2015,applebaum21, gutcke22, katz_impact_2024,brauer25, koutsouridou_nefertiti_2026, richardson_imf_2026, gjergo_massive_2026}. 

On top of this, a third major source of uncertainty arises from the treatment of stellar nucleosynthesis yields. Indeed, yield predictions strongly depend on the assumptions about stellar evolution, mass loss, rotation, convective mixing, and explosion mechanisms \citep{woosley95, meynet02,limongi18, nomoto13}, leading to substantial discrepancies among different sets of yields for both Pop~II/I and Pop~III stars \citep{iwamoto05, heger02, heger10,nomoto13}. These differences strongly affect the predicted abundance ratios of key elements, such as C, N, O, Fe, and Zn, and often translate into non-unique interpretations of observed chemical patterns \citep{romano10, romano19, romano20}. Nevertheless, most chemical-evolution models and hydrodynamical simulations adopt a single set of stellar yields per stellar population, rarely exploring or propagating the associated theoretical uncertainties into their predictions \citep[with notable exceptions, e.g.,][]{buck_challenge_2021}.

While the Pop~III pre-enrichment, the stochastic IMF sampling, and stellar yield uncertainties have each been recognized as important factors in shaping chemical evolution, their combined impact remains largely unexplored in standard modeling approaches. Their simultaneous treatment is particularly challenging, as it requires to keep track of discrete enrichment events while accounting for statistical population sampling of stars as well as for uncertainties in nucleosynthetic outputs. As a result, current models struggle to provide a unified and flexible framework capable of interpreting the chemical diversity observed across different environments, from the earliest galaxies to present-day metal-poor systems.
To address these challenges, we introduce \textsc{CRIMSONS}\footnote{(\CRIMSONS = Chemical evolution with the Random sampling of the Initial Mass function: Studying the Origin of Nucleosynthetic Stellar products)}, a new chemical evolution tool designed to self-consistently model the chemical enrichment under stochastic IMF sampling while explicitly accounting for Pop~III pre-enrichment and multiple sets of stellar yields. \CRIMSONS follows the evolution of individual stars drawn via Monte Carlo sampling of the IMF and tracks the enrichment of 30 chemical elements from H to Zn, including contributions from SNe, Type Ia SNe, and asymptotic giant branch (AGB) stars.

\CRIMSONS is designed as a flexible tool, able to provide both theoretical and observational researchers with a common framework for the study of chemical enrichment of different environments.
Thanks to its modular design, \CRIMSONS can serve in a broad variety of scientific applications. 
In this paper, we describe the underlying methodology of \CRIMSONS (Sect.~\ref{sec:model}), illustrate its capabilities through representative applications (Sect.~\ref{sec:results}),\addition{and discuss its implications for early galaxy evolution and low-metallicity systems} (Sect.~\ref{sec:conclusions}). \addition{ The first public online version of \CRIMSONS is available at \url{martina-rossi.it/crimsons.html} as web-based tool and downloadable via \texttt{pip} as Python package.}

%-------------------------------------------------------------------
%                           The CRIMSONS model
%-------------------------------------------------------------------

\section{The \CRIMSONS tool}
\label{sec:model}

\CRIMSONS is a chemical evolution tool designed to model the enrichment history of a single stellar population by explicitly accounting for stochastic sampling of the IMF, and allowing to treat separately the chemical feedback of Pop~III and Pop~II/I stars. \addition{In its current implementation, \CRIMSONS adopts a closed-box framework, neglecting gas inflows, outflows, and spatial mixing. This choice provides a controlled setup in which the effects of IMF sampling, stellar yields, and Pop III pre-enrichment on the resulting chemical patterns can be investigated independently of additional processes associated with gas flows and mixing.} In this section, we outline the fundamental assumptions and methodology adopted in \CRIMSONS\!\!: the stochastic IMF sampling technique (Sect.~\ref{sect:stochasticIMF}), the chemical evolution model (Sect.~\ref{sec:chemical_evolution}), the fiducial configuration for stellar yields  (Sect.~\ref{sec:fiducial perscription}), and the publicly available online implementation (Sect.~\ref{sec:public tool}).

%-------------------------------------------------------------------
%                   Stochastic IMF Sampling Procedure
%-------------------------------------------------------------------
\subsection{Stochastic IMF Sampling Procedure}
\label{sect:stochasticIMF}

\texttt{CRIMSONS} implements a Monte Carlo-based sampling of the IMF, a method widely adopted in previous works \citep[e.g.,][]{debennessuti17, rossi21,rossi24, pagnini23, koutsouridou23, koutsouridou_nefertiti_2026,  ciabattini25}. This method stochastically samples individual stellar masses according to the IMF probability distribution, producing a discrete representation of the stellar population.

The sampling procedure proceeds as follows \citep[see also][]{debennessuti17, rossi21}. 
As a preliminary step, we normalize the IMF, $\phi(m)$, and build the IMF's cumulative distribution function (CDF), $P(m)$. 
The IMF normalization is made over the mass range and with a fixed grid resolution of $0.1 \msun$. The mass range has a different extension depending on the chosen shape of the IMF, and whether a Pop~III or Pop~II/I stellar burst is selected (e.g., see our fiducial set up described in Sec.\ref{sec:fiducial perscription}). 
As a consequence of the IMF normalization, $P(m)$ is defined in $[0,1]$. Next, in order to determine the mass of each star, we extract a random number $r$ from an uniform distribution between $0$ and $1$, and invert the CDF, i.e. we solve $P(m)=r$. To do so, we search for the two consecutive CDF values between which $r$ falls, say $P(m_i)$ and $P(m_{i+1})$, and assign a stellar mass to $r$ as the mean value $m_{\star,i} = (m_i + m_{i+1})/2$. 
We repeat the extraction iteratively, until the cumulative mass of extracted stars equals the mass of the stellar burst, $M_\star$, that is when $M_\star=\sum_i m_{\star,i}N_{\star,i}$, where $N_{\star,i}$ is the number of extracted stars with mass $m_{\star,i}$. The resulting ensemble of stellar masses represents a discrete realization of the IMF, referred to as the {\it effective} IMF. The {\it effective} IMF is stored and can be directly compared to the theoretical input, allowing for ensemble statistics and diagnostic visualization. 

%-------------------------------------------------------------------
%                         Chemical Evolution
%-------------------------------------------------------------------
\subsection{Chemical evolution}
\label{sec:chemical_evolution}

\begin{table*}
\caption{Summary of stellar yield sets implemented in \CRIMSONS, grouped by enrichment channel and stellar population. The mass ranges correspond to the initial stellar mass of progenitors contributing to each channel.}
\label{tab:yields_summary}
\centering
\begin{tabular}{|l|p{3cm}|p{5.5cm}|p{5.5cm}|}
\hline
\textbf{Channel} & \textbf{Mass Range [$M_\odot$]} & \textbf{Yields Pop~II} & \textbf{Yields Pop~III} \\
\hline
AGB & 2–8 & \citet{vanDenHoek97}; \citet{nomoto13} (based on \citet{karakas10})& \citet{vanDenHoek97}; \citet{meynet02}; \citet{nomoto13} (based on \citet{campbell08}) 
\\
\hline
SNe & Pop~II: 13-120 (or 20-40 for NK); Pop~III: 10–100 (or 20-40 for NK) & \citet{limongi18}; \citet{nomoto13} & \citet{heger10}; \citet{iwamoto05}; \citet{nomoto13} \\
\hline
Type~Ia SNe & $<8$ & \citet{iwamoto99} & \citet{iwamoto99} model W70 \\
\hline
PISNe & 140–260 & -- & \citet{heger02}; \citet{nomoto13} \\
\hline
\end{tabular}
\end{table*}

\CRIMSONS follows the chemical enrichment of the ISM produced by a stellar population, accounting for the mass- and metallicity-dependent evolution of individual stars. The model assumes a single star formation event at the initial time $t_{0}$, where a total stellar mass $M_{\star}$ is formed out of a gas with initial metallicity $Z_{\rm ISM}\left(t_0\right)=Z_0$.
Stars are assumed to form with the same metallicity of the ISM, i.e. $Z_\star=Z_0$, and are classified as Pop~III if $Z_\star<Z_\mathrm{crit}=10^{-4.5}Z_\odot$ \citep{bromm01, schneider03, omukai05}, and Pop~II/I otherwise.
The chemical enrichment is modeled within a closed-box framework, where the gas inflows, outflows \addition{, and spatial mixing are neglected.}
These assumptions are appropriate for modeling  environments where the chemical evolution is dominated by a limited number of enrichment events, such as early star-forming systems, or where the chemical evolution is determined by localized enrichment episodes. In these cases, \CRIMSONS provides a framework to disentangle the effects of stellar evolution, stochastic IMF sampling, and nucleosynthesis.
At the same time, \CRIMSONS is conceived as a modular building block, whose output can be easily incorporated into more complex analytical models or numerical simulations.

As stars evolve, the composition of the ISM is updated based on stellar ejecta.
Stellar lifetimes for Pop~III stars are computed following  \citet{schaerer_properties_2002}, while for Pop~II/I we exploit relations from \cite{raiteri96}, which depend on both the initial stellar mass and metallicity. Note that differences between alternative stellar lifetime prescriptions are expected to have a negligible impact on our results, since in the case of single, instantaneous star formation bursts the chemical enrichment is driven by massive stars evolving on comparable short timescales.
At each time step, stars that reach the end of their life contribute to the chemical enrichment of the ISM through different nucleosynthesis channels: AGB winds, core-collapse SNe, SNe Ia, and, for Pop~III stars, PISNe.
We follow the evolution for 30 chemical elements individually, from hydrogen to zinc. 
Overall, the ISM chemical enrichment is followed for $13$ Gyr, and with a temporal resolution of $1$ Myr. 
Finally, since a single stochastic realization of the IMF is not statistically representative of any stellar population, in this work we present the results for a number of realizations $N_{\rm run}=50$. 
Note that the public tool allows the user to freely set $N_{\rm run}$, as well as $M_\star$ and $Z_0$.

\subsubsection{Chemical enrichment channels}
Here we detail the different enrichment channels contributing to the chemical evolution, as summarized in Table~\ref{tab:yields_summary}. \addition{For each chemical enrichment channel, we define a specific stellar mass range covered by the provided yield tables. For stars with masses within this range, the yields are computed via linear interpolation between the adjacent tabulated masses. For stellar masses falling outside this range, the chemical enrichment is computed by scaling the yields of the closest tabulated stellar mass linearly proportional to the stellar mass. 
Furthermore, we divide each enrichment channel into two types based on metallicity: Pop~III and Pop~II/I. For the Pop~III case, we assume that any star forming from gas below $Z_{\rm crit}$ uses zero-metallicity yields. For Pop~II/I stars, we linearly interpolate the chemical enrichment between tables of different metallicities. If the metallicity falls outside the range of the selected yield set, we assume the yields of the closest tabulated metallicity.}

\paragraph{Asymptotic Giant Branch (AGB):} Stars with initial masses in [2-8] $\mbox M_\odot$ enrich the ISM through stellar winds, primarily in carbon, nitrogen, and oxygen and other elements processed in stellar interiors. \CRIMSONS includes the yields from \cite{vanDenHoek97}, \cite{nomoto13}, and the rotating stellar models ($v = 300$ km/s) from \cite{meynet02}, for Pop~III stars, and those from \cite{vanDenHoek97} and \cite{nomoto13} for Pop~II/I stars. 
We note that yields for low- and intermediate-mass Pop~II/I stars from \citet{nomoto13} actually rest on the grids computed by \citet{karakas10}, while for Pop~III (Z=0) stars they are computed based on the yields (and remnant masses) from \citet{campbell08}.

\paragraph{Supernovae (SNe):} Stars with initial masses $\gtrsim10~\mbox M_\odot$ are assumed to end their lives as SNe, which represent a key channel for the chemical enrichment and whose yields heavily depend on the progenitor mass, rotation rate, and the explosion mechanism. In \CRIMSONS \!, Pop~III and Pop~II/I are treated separately, allowing for different and independent configurations of yield sets and explosion energies ($\rm E_{SN}$).

For Pop~III stars in the mass range $[10-100]~\mbox M_\odot$, \CRIMSONS allows for different choices. 
The most flexible set of yields is from \cite{heger10} as it includes specific models for faint (Faint, $\rm E_{SN}=[0.3-0.6] \times 10^{51}$~erg), core-collapse SNe (CC, $\rm E_{SN}=[1.2-1.5] \times 10^{51}$~erg), high-energy (High, $\rm E_{SN}=[1.8-3.0] \times 10^{51}$~erg), and hypernovae (Hyper, $\rm E_{SN}=[5.0-10.0] \times 10^{51}$~erg) SNe. 
For this set of stellar yields, \CRIMSONS allows to select a specific explosion type, or to assume a energy distribution function (see Sect. \ref{sec:EDF}), or to randomly assign the explosion energy assuming a constant probability (i.e. $25\%$ for each type). We refer to the latter case as the \textit{all} SNe model.
Additional sets of Pop~III SNe yields implemented in the code are those from \citet{iwamoto05} and \citet{nomoto13}. In the first one, the SNe explosion energy scales with the progenitor mass according to the linear relation $E_{\rm SN}=0.79\times10^{51}(M_\star/25 M_{\odot}){\rm\, erg}$. In the latter, two different prescriptions for $E_{\rm SN}$ are provided: one corresponding to standard CCSNe, spanning the $[11-100]\mbox M_\odot$ mass range with fixed explosion energy of $E_{\rm SN}=10^{51}{\rm\, erg}$, and another for hypernovae covering the mass range $[20-100]\mbox M_\odot$ with $E_{\rm SN}\in\left[1.0-7.1\right]\times10^{52}{\rm\, erg}$ depending on the progenitor mass.

For Pop~II/I stars with initial masses in $\left[13-100\right] M_{\odot}$, or $[20-40]M_{\odot}$, depending on the adopted stellar yields, \CRIMSONS includes the sets of yields from \cite{limongi18} and \cite{nomoto13}.
The \citet{limongi18} yields account for rotationally–enhanced mixing in stellar evolution and nucleosynthesis, providing models for non–rotating ($v=0{\rm\, km/s}$) %, moderate-rotating ($v=150{\rm\, km/s}$) 
%\textcolor{magenta}{[DR: $v=150{\rm\, km/s}$ already means stars are fast rotators, moderate is 20--60, but it's not included in Marco's study]}, 
and fast-rotating ($v=150{\rm\, km/s}$, $v=300{\rm\, km/s}$) stars, all computed assuming a fixed explosion energy of $E_{\rm SN}=10^{51}{\rm\, erg}$.
The set of yields from \cite{nomoto13} distinguishes between CCSNe and hypernovae, as described above for Pop~III stars.

\paragraph{Type Ia Supernovae (SNe Ia):} SNe Ia provide a dominant contribution to iron and iron-peak elements at late times and play a key role in shaping the decline of [$\alpha$/Fe] abundance ratios observed in stellar populations. 
In \CRIMSONS we offer two alternative prescriptions to model the enrichment from SNe Ia. In the first, the user can explicitly set both the timing and the amplitude of the SNe Ia contribution through two input parameters: (i) the delay time at which all SNe Ia explode ($\Delta t_{\rm SNIa}$), and (ii) the mass fraction of stars with initial mass $m_{\star}<8M_\odot$ that explode as Type Ia supernovae ($f_{\rm SN~Ia}$).
Alternatively, the user may adopt a delay–time distribution (DTD) formalism, implemented here in two versions: one from \citet{matteucci06} and calibrated on MW observations, the other from \citet{vogelsberger13} and calibrated on SLOAN II observations \citep{maoz12}.
% Alternatively, the user may adopt a delay–time distribution (DTD) formalism, which is implemented in the versions of \citet{matteucci06}, calibrated on MW observations, or of \citet{vogelsberger13}, calibrated on SLOAN II observations \citep{maoz12}.
The first approach offers maximal flexibility, and for instance it is particularly useful for exploring the impact of SNe Ia in a Pop~III enrichment scenarios. On the contrary, the DTD formalism is generally more appropriate for Pop~II/I chemical evolution studies.
For SNe Ia \CRIMSONS relies on the yields from \cite{iwamoto99}, in which several models are provided (W7, W70, WDD1, WDD2, WDD3, CDD1, and CDD2, see the original publication for further details). In the case of Pop~III progenitors, the W70 model is recommended due to its compatibility with metal-free conditions.

\paragraph{Pair-Instability SNe:} PISNe can be included in \CRIMSONS for Pop~III stars with initial masses in the range $\left[140-260\right]{\rm\, M_\odot}$. These extremely energetic explosions contribute with significant amounts of both $\alpha$-elements and iron-peak elements to the ISM, and are particularly relevant for early metal enrichment. The implemented set of stellar yields is that from \cite{heger02} and \cite{nomoto13}. Finally, \addition{PISN contribution can be excluded, for example with the "no PISN" option in the web interface of \CRIMSONS.}

%-------------------------------------------------------------------
%                           The Online tool
%-------------------------------------------------------------------
\subsection{The \addition{public} tool}
\label{sec:public tool}

\CRIMSONS is a flexible and user-friendly tool designed to be applicable across a wide variety of astrophysical environments, from primordial metal-poor systems to more evolved and chemically enriched ones. Its modular structure allows for a systematic exploration of the influence of different physical assumptions on the chemical evolution of galaxies and stellar populations.
For this reason, we have released \CRIMSONS \addition{in two formats: as an interactive web-based tool and a Python package}, enabling the broader community to fully exploit its capabilities. The \CRIMSONS \addition{web-tool} is available at \url{martina-rossi.it/crimsons.html}, together with some python-based scripts as examples to interact with the outputs. \addition{The \CRIMSONS Python package is downloadable via \texttt{pip} and the documentation can be found at \url{crimsons.readthedocs.io}.} 
In the following section, we outline the major features of \addition{both \CRIMSONS implementations; for further details, we refer the reader to the user guides on their respective website}. 

At first, along with a few other inputs (see the official user guide for details), the tool requires the initial gas metallicity $Z_0$, upon which it automatically selects the stellar population: Pop~III stars if $Z_0\leq10^{-4.5} Z_{\odot}$, Pop~II/I stars otherwise.
Next, the tool enables the IMF selection among (i) a Larson-like IMF with free $m_{\rm ch}$ for Pop~III, and fixed $m_{\rm ch}=0.35 M_\odot$ for Pop~II \addition{(\Cref{eq:larson})}, (ii) a power-law form ($\phi(m_\star)\propto m_\star^{-\alpha_s}$) with free slope $\alpha_s$ for Pop~III, and fixed Salpeter-like $\alpha_s=2.35$ for Pop~II/I stars, and (iii) a flat IMF in the case of Pop~III stars, or a Kroupa IMF \citep{kroupa03} for Pop~II/I stars \addition{ defined as $\phi(m_\star)\propto m_\star^{-\alpha_k}$, with $\alpha_k = 1.3$ for $m_\star < 0.5 \msun$ and $\alpha_k = 2.3$ otherwise}.

Along with the IMF, sets of yields can be selected among those presented in Sec. \ref{sec:chemical_evolution} and summarized in Table \ref{tab:yields_summary}. \addition{When a selected yield set features additional parameters, such as stellar rotational velocities \citep[e.g.,][]{limongi18}, internal mixing and explosion energy \citep[e.g.,][]{heger10}, or specific explosion models \citep[e.g.,][]{iwamoto05}, the user can either select a single, fixed parameter value or sample them from a user-defined probability distribution (see also Sect.~\ref{sec:EDF})}.

We plan on adding more yield sets in the future and we refer to the user guide on the website for the updated list of yields, as well as for any other major update of the model. 

%-------------------------------------------------------------------
%                           The fiducial persciption
%-------------------------------------------------------------------
\subsection{Fiducial prescriptions}
\label{sec:fiducial perscription}

\begin{table*}
\centering
\caption{Parameters of our Pop~III and Pop~II/I fiducial model.}
\label{tab:fiducial_parameters}
\begin{tabular}{|l|l|c|c|}
\hline
\textbf{Component} & \textbf{Parameter} & \textbf{Pop~III} & \textbf{Pop~II} \\
\hline
\multirow{3}{*}{IMF} 
    & Functional form & Larson IMF & Larson IMF \\
    & $m_{\rm ch}$ & $10~M_\odot$ & $0.35~M_\odot$ \\
    & Mass range & $0.8$--$1000~M_\odot$ & $0.1$--$100~M_\odot$ \\
\hline
\multirow{2}{*}{SNe} 
    & Yields & \citet{heger10} & \citet{limongi18}, $v = 0$ km/s \\
    & Explosion types & All type & Standard CCSNe \\
\hline
PISN
    & Yields & \citet{heger02} & -- \\
\hline
AGB stars 
    & Yields & \citet{meynet02} (rotating) & \citet{vanDenHoek97} \\
\hline
\multirow{2}{*}{SNe Ia} 
    & Yields & \citet{iwamoto99} (W70) & \citet{iwamoto99} (W7) \\
    & $\Delta t_{\rm SN Ia}$ & $[30-60]$ Myr & $[30-60]$ Myr \\
\hline
\end{tabular}
\end{table*}
Among the many possible combinations of parameter offered by \texttt{CRIMSONS}, it is useful to define a reference configuration that can serve both as a baseline for comparisons and as a physically motivated starting point for further exploration. For this purpose, we adopt a fiducial set of prescriptions for the IMF and stellar yields for both Pop~III and Pop~II/I stellar populations. These choices are motivated by previous studies and constrained through comparisons with observational data of metal-poor stellar populations. A summary of the adopted parameters is provided in Table~\ref{tab:fiducial_parameters}. 

In our Pop~III fiducial model, the mass distribution is distributed in the range $m_{\star} = \left[0.8-1000\right] M_{\odot}$ according to a Larson IMF \citep{larson98}:
\begin{equation} \label{eq:larson}
  \phi(m_\star) = \frac{dN}{dm_{\star}}\propto m_\star^{-2.35}\exp(-m_\mathrm{ch}/m_\star) ,
\end{equation}
where $m_{\rm ch}=10{\rm\,M_\odot}$ is the characteristic mass. This choice reflects the expectation of a top-heavy IMF for primordial star formation and is consistent with both theoretical predictions and observationally motivated models \citep{rossi21, pagnini23, koutsouridou23, rossi23, rusta_pristine_2026}. 
For the nucleosynthesis prescriptions we adopt rotating models ($v = 300$ km/s) from \citet{meynet02} for AGB stars, supernova yields from \citet{heger10} for massive stars, and pair-instability supernova yields from \citet{heger02}. \addition{\citet{heger10} provides yields for different explosion energies and internal mixing. For the former, we adopt an energy distribution function (see Sect. \ref{sec:EDF}), for the latter, we adopt yields with mixing parameter between $0.039$ and $0.158$, consistent with results of the chemical abundances of C-enhanced metal-poor stars in the Galactic halo \citep{koutsouridou23}.} Type Ia supernovae are not included in the fiducial Pop~III configuration ($f_{\rm SN~Ia}=0$), since their contribution is expected to be negligible during the earliest phases of chemical enrichment, \addition{when it is driven exclusively by massive star progenitors. At later times, when delayed Pop~III SNe Ia would theoretically explode, the overall chemical evolution is heavily dominated by Pop~II/I star formation (which includes SNe Ia), rendering any residual Pop~III SN~Ia contribution negligible} \citep{salvadori09}.

For Pop~II/I stars, the fiducial model adopts a Larson-type IMF with $m_{\rm ch}=0.35\,M_{\odot}$ over the mass range $[0.1-100]~M_\odot$. The adopted stellar yields are those from \citet{vanDenHoek97} for AGB stars, the non-rotating ($v=0{\rm\, km / s}$) models of \citet{limongi18} for core-collapse supernovae, and the W7 prescription from \citet{iwamoto99} for Type Ia supernovae. This configuration has been shown to reproduce key features of the chemical evolution of the Galactic halo \citep{koutsouridou23,rossi24_1} and nearby ultra-faint dwarf galaxies \citep{rossi24}.
\addition{We note that chemical enrichment from PISN is not included in the present analysis. Within the adopted closed-box framework, their large metal yields can dominate the resulting enrichment pattern, potentially masking the differences among the stellar yield prescriptions that we aim to isolate and quantify}.
\addition{We emphasize that adopting these specific fiducial prescriptions serves as a calibrated baseline grounded in previous literature. While those studies constrained these parameters using more complex models against observations, \CRIMSONS adopts them as physical inputs to establish a reference configuration. From this baseline, we explore the impact of alternative yield sets, explosion energies, and IMF parameters throughout the rest of this work.}

\subsection{Energy distribution function (EDF)}\label{sec:EDF}

An additional degree of freedom that impacts the resulting chemical enrichment is the distribution of supernova explosion energies. This is described through the Energy Distribution Function (EDF), which regulates the relative probability of different supernova explosion energies.  
For Pop~III stars in the mass range $m_\star = 10$–$100\,M_\odot$, the explosion energy is expected to span a wide range of values depending on the explosion mechanism, stellar rotation, and internal structure of the progenitor. However, current theoretical models do not provide robust constraints on the statistical distribution of these energies. Observationally inferred explosion energies span nearly two orders of magnitude, but the underlying probability distribution remains essentially unknown.

To account for this uncertainty, \texttt{CRIMSONS} implements an EDF that regulates the relative probability of different supernova explosion energies. Following the approach adopted in \citet{koutsouridou23}, we assume a mass-independent EDF as
\begin{equation}
\frac{dN}{dE_\star} \propto E_\star^{-\beta},
\end{equation}
where $E_\star$ is the explosion energy and $\beta$ is a free parameter. Based on this distribution, an explosion energy is assigned to each Pop~III SNe progenitor in the mass range $10$–$100\,M_\odot$.
Note that, by varying the parameter $\beta$ in the EDF one can preferentially weight low-energy (e.g. faint SNe) or high-energy explosions (e.g. hypernovae), allowing different physical scenarios for Pop~III explosions to be explored. 

\begin{figure}
    \centering
    \includegraphics[width=\columnwidth]{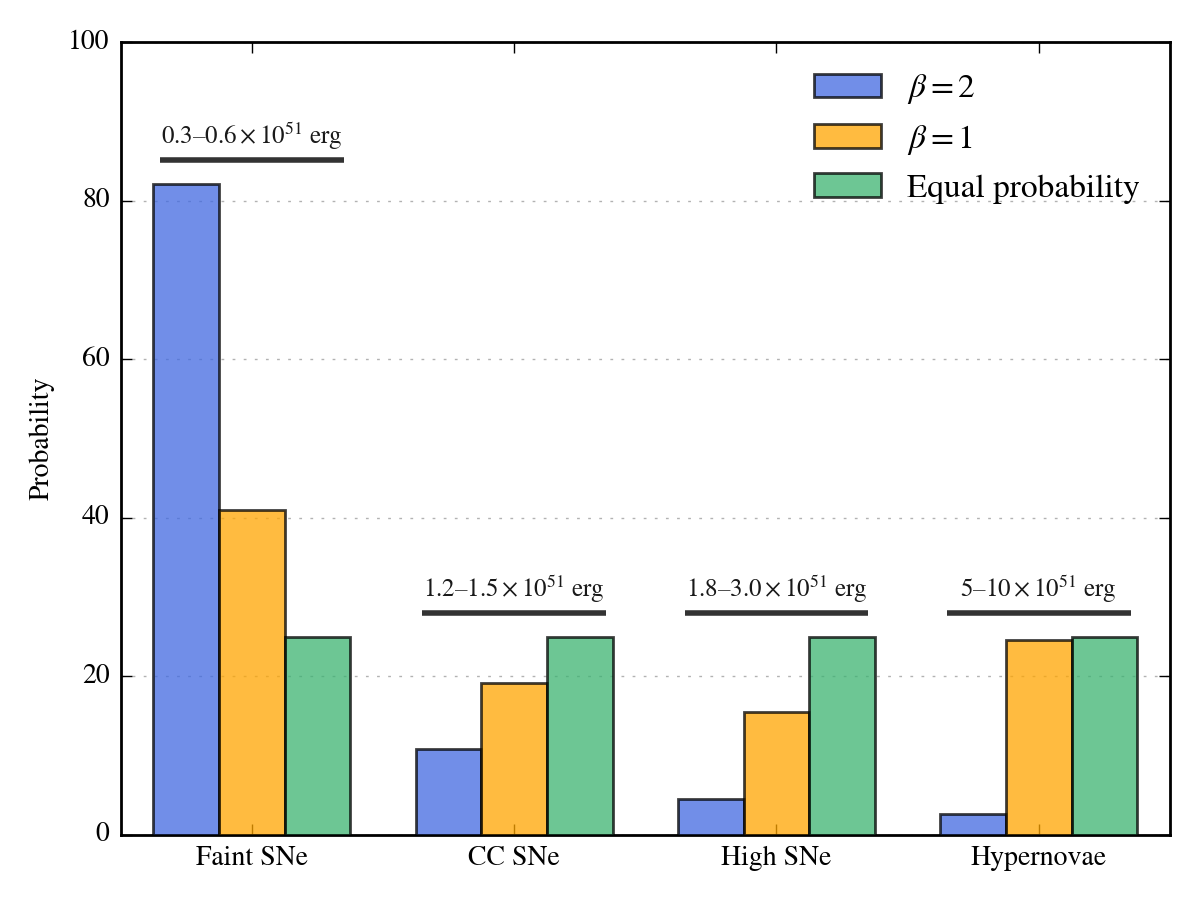}
    \caption{Integrated probability for Pop III massive stars to explode as Faint, CC, High, or Hyper SNe, for different values of $\beta$. The yellow histogram corresponds to $\beta = 1$, the blue histogram to $\beta = 2$, while the green histogram represents the reference case in which all Pop III explosion energies are equally probable.}
    \label{fig:edf}
\end{figure}

In Figure~\ref{fig:edf} we show the integrated probability for Pop III massive stars to explode as Faint, CC, High, or Hyper SNe, for different values of $\beta$. 
We see that a larger $\beta$ strongly favors low-energy explosions, resulting in a higher probability for faint SNe with respect to more energetic types. Conversely, smaller values of $\beta$ produce more comparable probabilities for the different explosion channels. For reference, we also show the case of equal probability (25\% each) for all explosion types, which is the adopted configuration assumed in the examples presented in this work. 
This choice is purely illustrative, it is used here only to provide a neutral baseline configuration that allows different explosion models to be explored simultaneously, and it should not be interpreted as a physical prediction for the true distribution of Pop~III supernova energies. Note that for PISNe, i.e. Pop~III stars with masses in $[140, 260] M_{\odot}$, the explosion energy depends on the progenitor mass and is therefore not incorporated into the EDF formalism. For the adopted yield set, the ejecta yields scale approximately linearly with the explosion energy.

In realistic astrophysical environments, the distribution of SNe explosion energies is expected to depend on several factors, including the properties of the stellar population, the metallicity of the gas, and the physical conditions of the host system. For this reason, the EDF is implemented in \texttt{CRIMSONS} as a flexible and customizable component of the model. The parameter $\beta$ can be freely selected by the user, allowing the tool to tailor the EDF to the specific astrophysical environment being simulated, and to investigate how different assumptions on the Pop~III supernova population affect the resulting chemical enrichment.

%===================================================================
%                               RESULTS
%===================================================================
\section{Results} \label{sec:results}
%%%%%%%%%%%%%%%%%%%%%%%%%%%%%%%%%%%%%%%%%%%%%%%%%%%%%%%%%%%%%%
In this section, we present the results obtained with \texttt{CRIMSONS}, using simplified and controlled star-forming environments as proof-of-concept applications of the framework.
We quantify the effects of stochastic IMF sampling, compare the impact of different stellar yield sets, and assess the chemical signatures of Pop~III pre-enrichment on subsequent Pop~II/I populations. Together, these results show how the discrete nature of the  star formation process, the uncertainties on nucleosynthesis yields, and the early enrichment, shape the chemical evolution.

\subsection{IMF random sampling}
%%%%%%%%%%%%%%%%%%%%%%%%%%%%%%%%%%%%%%%%%%%%%%%%%%%%%%%%%%%%%%

%~~~~~~~~~~~~~~~~~~~~~~~~~~~~~~~~~~~~~~~~~~~~~~~~~~~~~~~~~~~

\begin{figure*}
    \centering
    \includegraphics[width=0.8\textwidth]{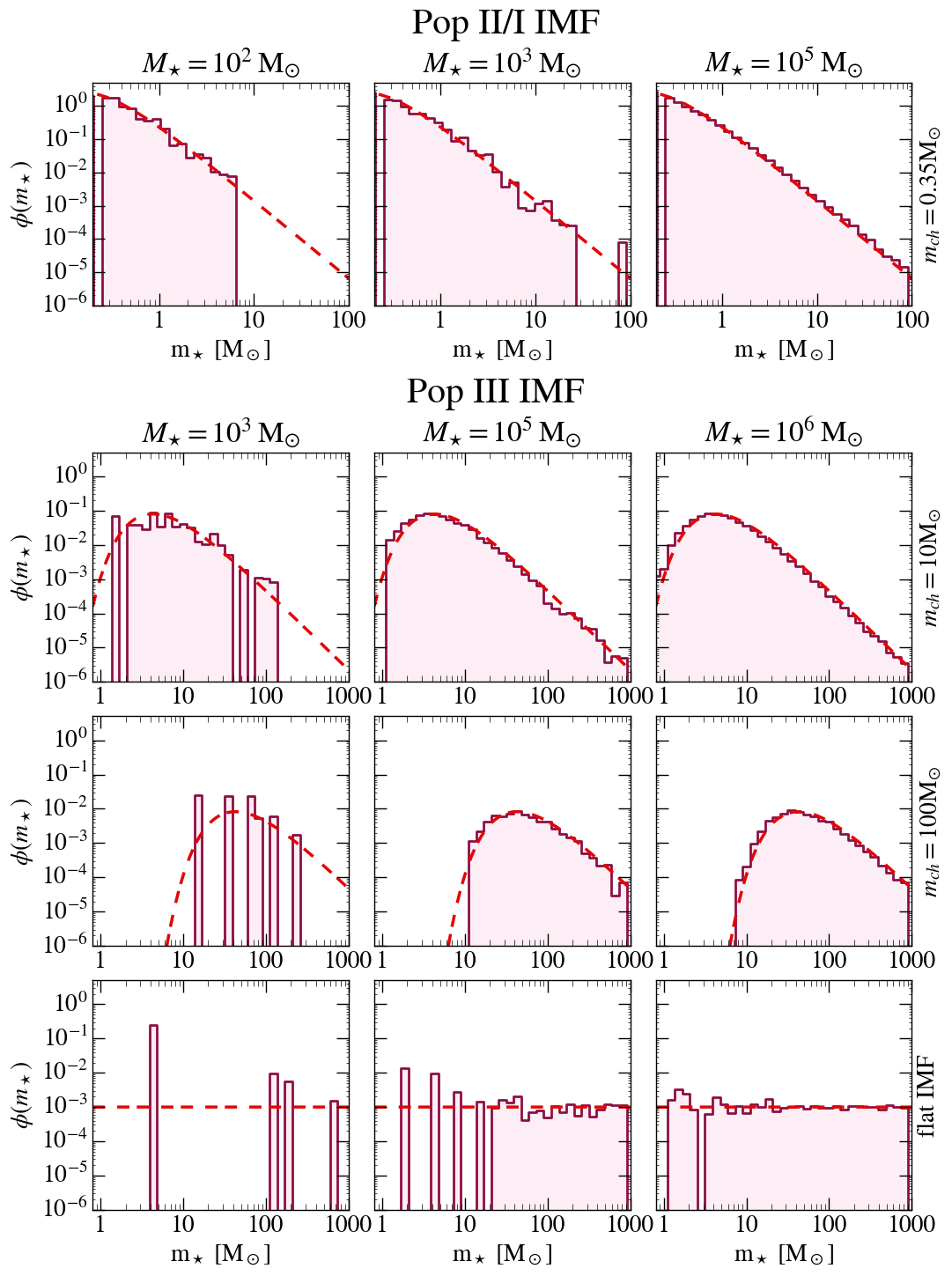}
    \caption{Comparison between the normalized analytic (dashed lines) and the effective IMF (histogram) resulting from our sampling procedure, for increasing stellar burst masses (from left to right). Each row shows a different IMF configuration, whether for a different characteristic mass, $m_{ch}$, for a Larson profile (top three rows) or a flat IMF (bottom row). \addition{The fiducial models for Pop~II/I and Pop~III are shown in the first two rows, respectively.}}
    \label{fig:figure1}
\end{figure*}

\Cref{fig:figure1} illustrates the effects of our random sampling procedure on the IMF of individual star formation bursts. In each panel, theoretical (dashed lines) and effective (histograms) IMFs are compared, for various total stellar mass formed (columns) and functional forms of the IMF (rows). 
We show separately the case of Pop~II/I (three panels on top) and Pop~III (lower panels) stellar bursts. 
Note that in these two scenarios both the mass range of the IMF and the total stellar mass formed in each burst ($M_\star$) might differ.

From \Cref{fig:figure1}, it is evident that the convergence between the theoretical and \textit{effective} IMF depends on both $M_{\star}$ and the underlying IMF shape. As $M_\star$ increases, any theoretical IMF becomes progressively better sampled, and the $M_\star$ values at which convergence is achieved depend on $m_{\rm ch}$. For a PopII/I-like IMF with $m_{\rm ch} = 0.35M_\odot$, convergence is already achieved at $M_\star \sim 10^5 M_\odot$, while for PopIII-like IMFs, with either a top-heavy form ($m_{\rm ch} \geq 10M_\odot$) or a flat slope, sampling completeness requires significantly larger stellar masses, typically $M_\star \gtrsim 10^5 - 10^6M_\odot$. 
On the contrary, at lower $M_\star$ the stochastic effects are enhanced, and for $M_\star \lesssim 10^5~M_\odot$ the number of formed stars is insufficient to properly sample any IMF shape. In these cases, the \textit{effective} IMF can significantly deviate from the theoretical expectation, especially in the high-mass regime.
Overall, these results indicate that the commonly adopted assumption of a fully sampled IMF only holds above a minimum stellar mass threshold, whose value depends on the IMF shape. Bottom-heavy IMFs, such as those typically used for Pop~II/I stars, reach completeness at $M_\star \sim 10^4$--$10^5\,M_\odot$, whereas Pop~III IMFs require much higher stellar masses to achieve a comparably well-sampled population, $M_\star\gtrsim 10^{6}\,M_\odot$.

%%%%%%%%%%%%%%%%%%%%%%%%%%%%%%%%%%%%%%%%%%%%%%%%%%%%%%%%%%%%%%
\subsection{Impact of the IMF sampling on the chemical enrichment}\label{sec:IMF_sampling_chemical_enrichment}
%%%%%%%%%%%%%%%%%%%%%%%%%%%%%%%%%%%%%%%%%%%%%%%%%%%%%%%%%%%%%%

\begin{figure*}
    \centering
    \includegraphics[width=0.8\textwidth]{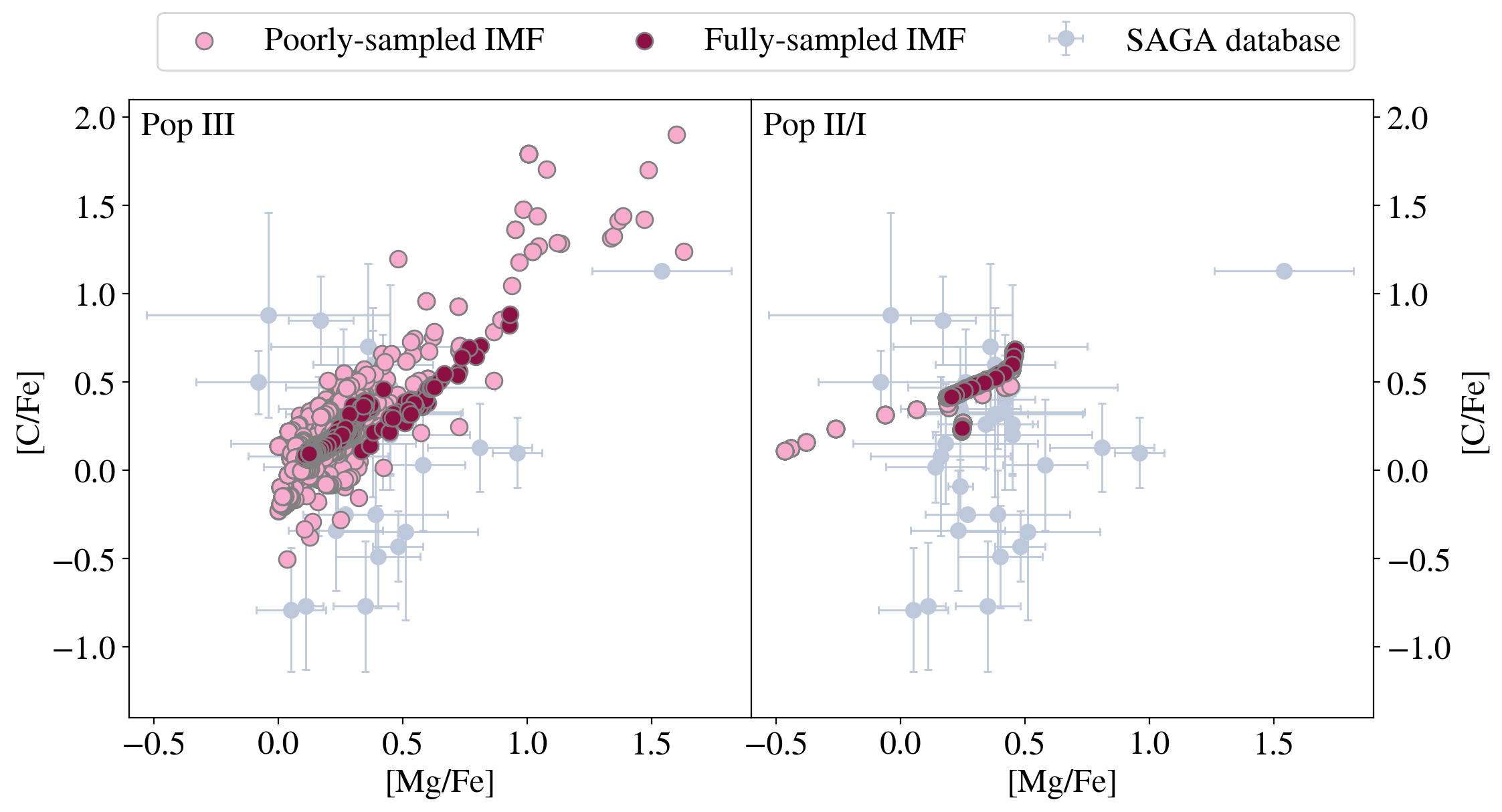}
    \caption{Diagnostic map of [C/Fe] versus [Mg/Fe] for Pop~III (left) and Pop~II/I (right) enrichment predicted by our fiducial model. For each \addition{of the 50} realizations, colored points show the gas-phase abundance at different times within the first $30$ Myr after the stellar burst \addition{with a resolution of 1 Myr}. Pink points correspond to scenario where the IMF is poorly sampled, while dark red points represent fully sampled IMF conditions. Gray points indicate observed abundances of stars in UFDs from the SAGA database \citep{saga_database}, corrected for carbon depletion following \citet{placco14}.}
    \label{fig:figure_2}
\end{figure*}

\begin{figure*}
    \centering
    \includegraphics[width=0.75\textwidth]{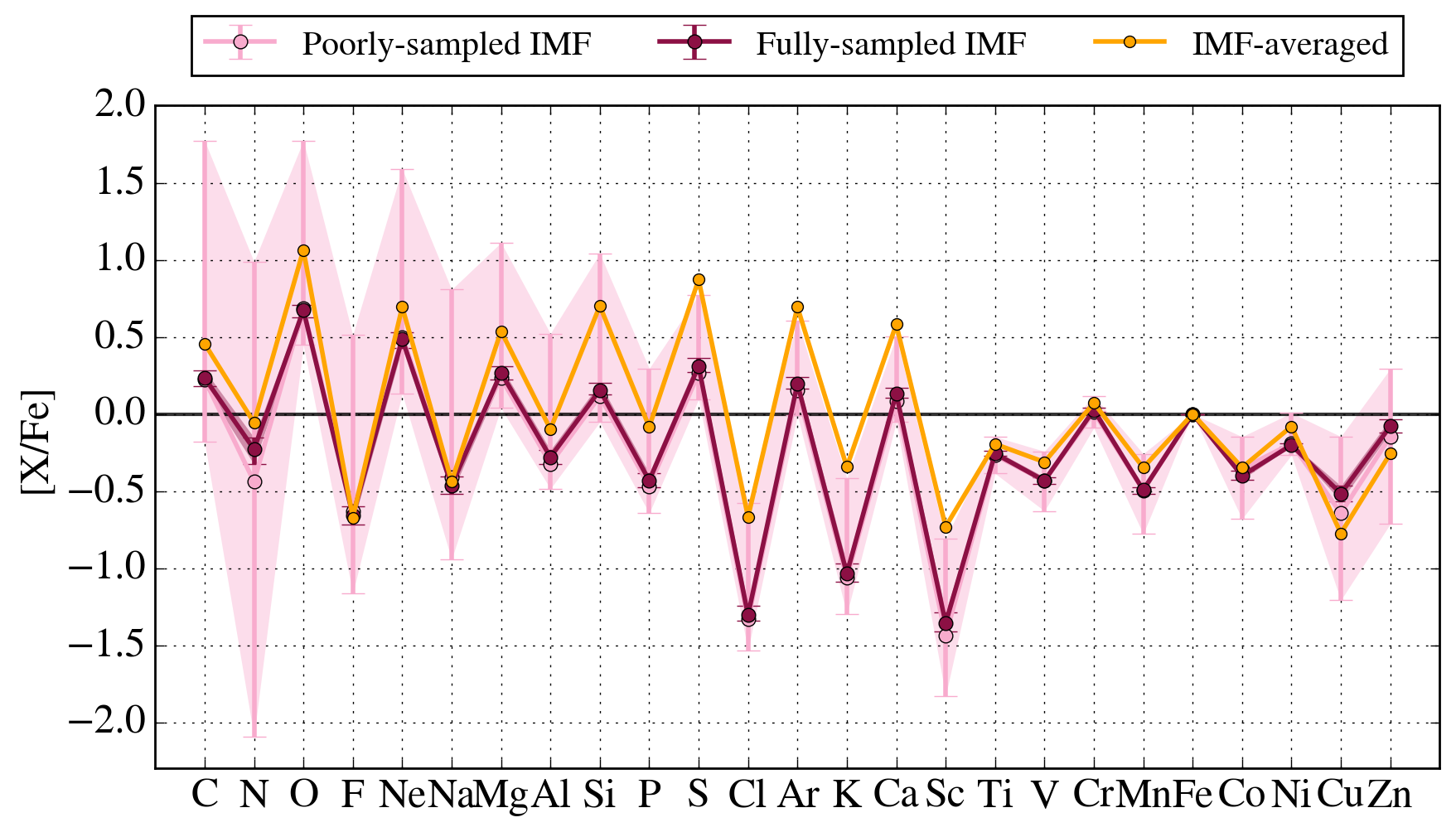}
    \caption{Comparison of the ISM abundance patterns 4 Myr after a Pop~III stellar burst for three cases: a poorly sampled IMF (pink), a fully sampled IMF (dark red), and the corresponding analytical averaged-IMF prediction (orange). \addition{For the sampled IMF, error bars and shaded regions indicate the 2.5/97.5 percentiles and solid lines the median over 50 realizations.}}
    \label{fig:figure_3}
\end{figure*}

We now investigate the impact of IMF sampling on the chemical enrichment of the surrounding ISM. 
\Cref{fig:figure_2} shows the predicted gas-phase [C/Fe] and [Mg/Fe] produced by a single burst of Pop~III (left panel) and Pop~II/I (right) stars. 
We assume here a poorly (pink, $M_{\star}=10^{3}M_{\odot}$ for Pop~III and $M_{\star}=10^{2}M_{\odot}$ for Pop~II) and a fully (dark red, $M_{\star}=10^{5}~M_\odot$) sampled IMF, and perform 50 different realizations.
In particular, each colored point represents the ISM composition at a time between $t_0$ (when the burst occurs) and $t_0+\Delta t$, where we set $\Delta t=30$ Myr and use a 1 Myr time resolution (see \Cref{app:time_sampling} for the time evolution of [C/Fe] and [Mg/Fe]).
Gray points are observed abundance ratios for stars in UFDs, taken from the SAGA database \citep{saga_database} and corrected for evolutionary carbon depletion using the prescription by \citet{placco14}. Observational data in these plots are shown only as a reference, thus they are not meant for a quantitative comparison, since such a comparison requires a full chemical-evolution model, whereas here we only track the ISM enrichment produced by a single star-formation event.

Inspecting \Cref{fig:figure_2}, we can see that a clear distinction emerges between the fully and poorly sampled cases. At low $M_\star$, the stochastic nature of the IMF leads to a large scatter in abundance ratios, driven by the random presence (or absence) of massive stars. In this regime, the chemical enrichment is controlled by very few SNe progenitors, making the final ISM abundances highly sensitive to the specific stellar masses sampled in each realization. The effect is particularly strong in the poorly sampled Pop~III case, where [C/Fe] and [Mg/Fe] span over 2 and 1.5 dex, respectively. Such large scatter is not driven by the use of multiple explosion energies in the adopted yields, as indeed comparable level of scatter arises even when the SN energy is fixed (e.g., faint SNe only), as shown in Appendix \ref{app:HW}.
In the Pop~II/I case, the scatter is less pronounced, although still significant at low $M_\star$, especially in [Mg/Fe]. 
Two factors contribute to this behavior: 
first, the adopted Pop~II/I yields show a relatively weak dependence on the progenitor mass. Hence, in both the poorly and the fully sampled case they produce a more homogeneous enrichment with respect to the Pop III scenario.
Moreover, in the poorly sampled case the random sampling of a bottom-heavy IMF often leads to realizations where the number of massive stars is very low (or even null), yielding a lower (if not negligible) and more scattered [Mg/Fe] than in the fully case.
% Moreover, in the poorly sampled case the random sampling of a bottom-heavy IMF often leads to realizations without massive stars, yielding a negligible SNe enrichment or more likely lower mass SNe which lead to  a lower and more scattered [Mg/Fe] than in the fully case.
%
When qualitatively compared with observations, these results show that the spread in [C/Fe] and [Mg/Fe] observed in UFD stars can naturally be explained with the stochastic nature of star formation at low $M_\star$ formed. 

Stochastic sampling effects also emerge when the analysis is extended to the whole set of abundance ratios, \addition{that are [X/Fe]}. \Cref{fig:figure_3} compares the gas-phase abundance ratios of all elements from carbon to zinc, in the metal-enriched regime set by Pop~III ejecta, prior to the onset of Pop~II star formation. The results are shown for a poorly sampled, a fully sampled, and the corresponding enrichment predicted using IMF-weighted yields. \addition{In all three cases of chemical enrichment we use the fiducial set of yields, which we recall are the \citet{heger10} for Pop~III stars.} The two sampled cases yield a comparable mean abundance ratios as expected when averaging over 50 realizations. What truly distinguishes them is the scatter: in the poorly sampled IMF, the enrichment can vary significantly from one realization to another, producing the much larger scatter seen across the 50 runs.
By contrast, the IMF-averaged enrichment deviates systematically from these mean values. \addition{As shown in the figure, this discrepancy is highly dependent on the specific element. The offset is most pronounced for the $\alpha$-elements (e.g., Si and Ca), where the IMF-averaged prediction overestimates the sampled means by more than a 0.5 dex. Conversely, other groups, such as odd-Z elements (e.g., Na, Al) and the iron-peak elements (e.g., Cr, Mn, Co, Ni), show virtually no systematic discrepancy between the IMF-averaged predictions and the sampled means.}

The differences can be traced back to how the IMF is handled in each model. In the sampled IMF cases, each mass bin contains an integer number of stars, so the mass distribution remains inherently discrete even when the IMF is fully sampled. In contrast, the IMF-averaged model assumes a continuous distribution with non-physical fractional star counts in each bin, which artificially smooths the contribution of massive stars.
This distinction is most significant at early times, when enrichment is dominated by massive SNe. For example, after $\sim15$ Myr (i.e., only 10 Myr later than in Figure \ref{fig:figure_3}), lower-mass SNe begin to dominate, and the offset between the sampled and IMF-averaged models decreases (see App. \ref{app:enrichment_15}). Still, a significant scatter remain in place for elements largely produced by low- and intermediate-mass stars, such as C and N.

%~~~~~~~~~~~~~~~~~~~~~~~~~~~~~~~~~~~~~~~~~~~~~~~~~~~~~~~~~~~

%---------------------- PopIII- yields ----------------------

\subsection{Impact of different stellar yields}

\begin{figure*}[h]
  \begin{subfigure}
  \centering
    \includegraphics[width=0.5\linewidth]{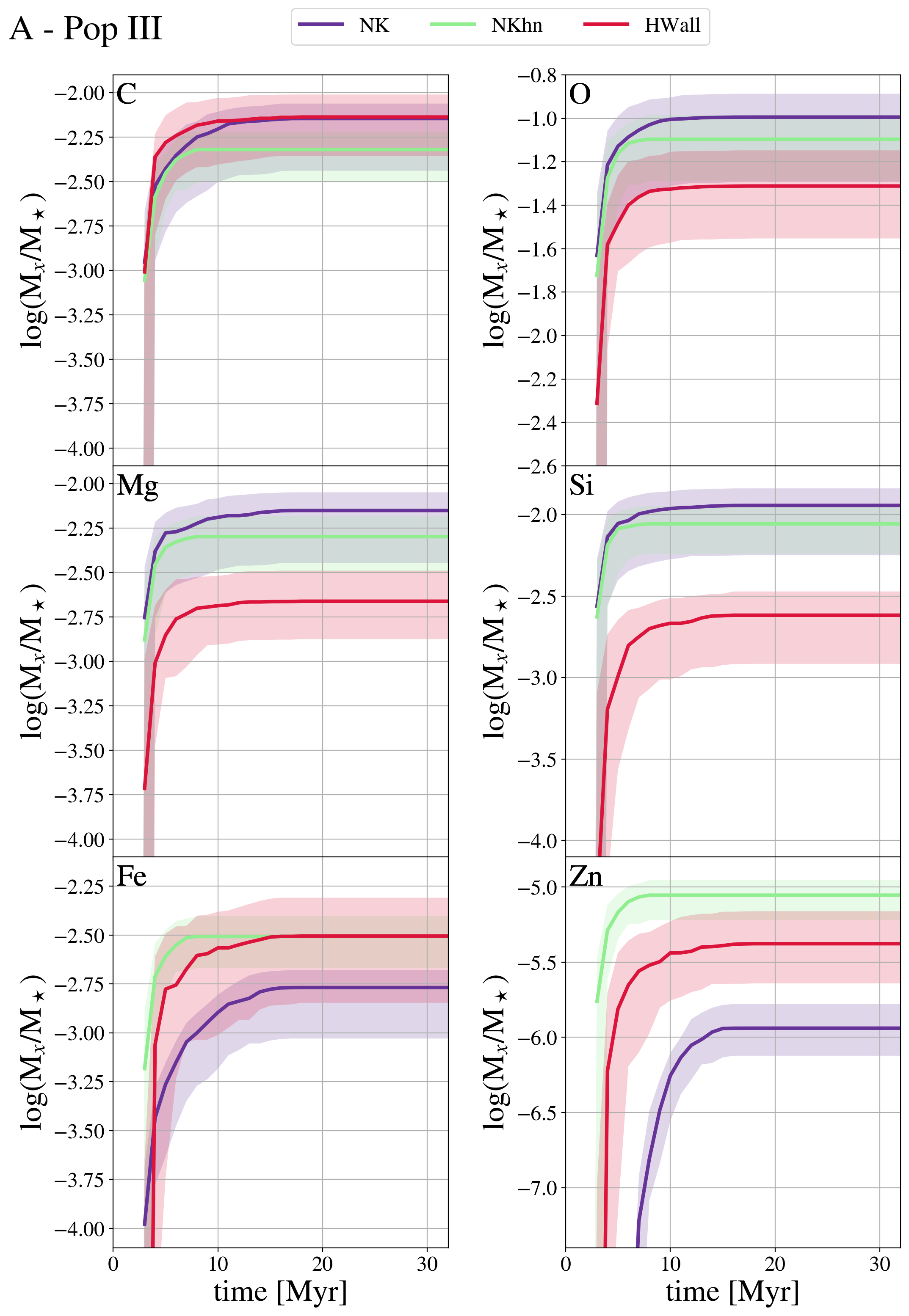}
  \end{subfigure}
  \begin{subfigure}
  \centering
    \includegraphics[width=0.5\linewidth]{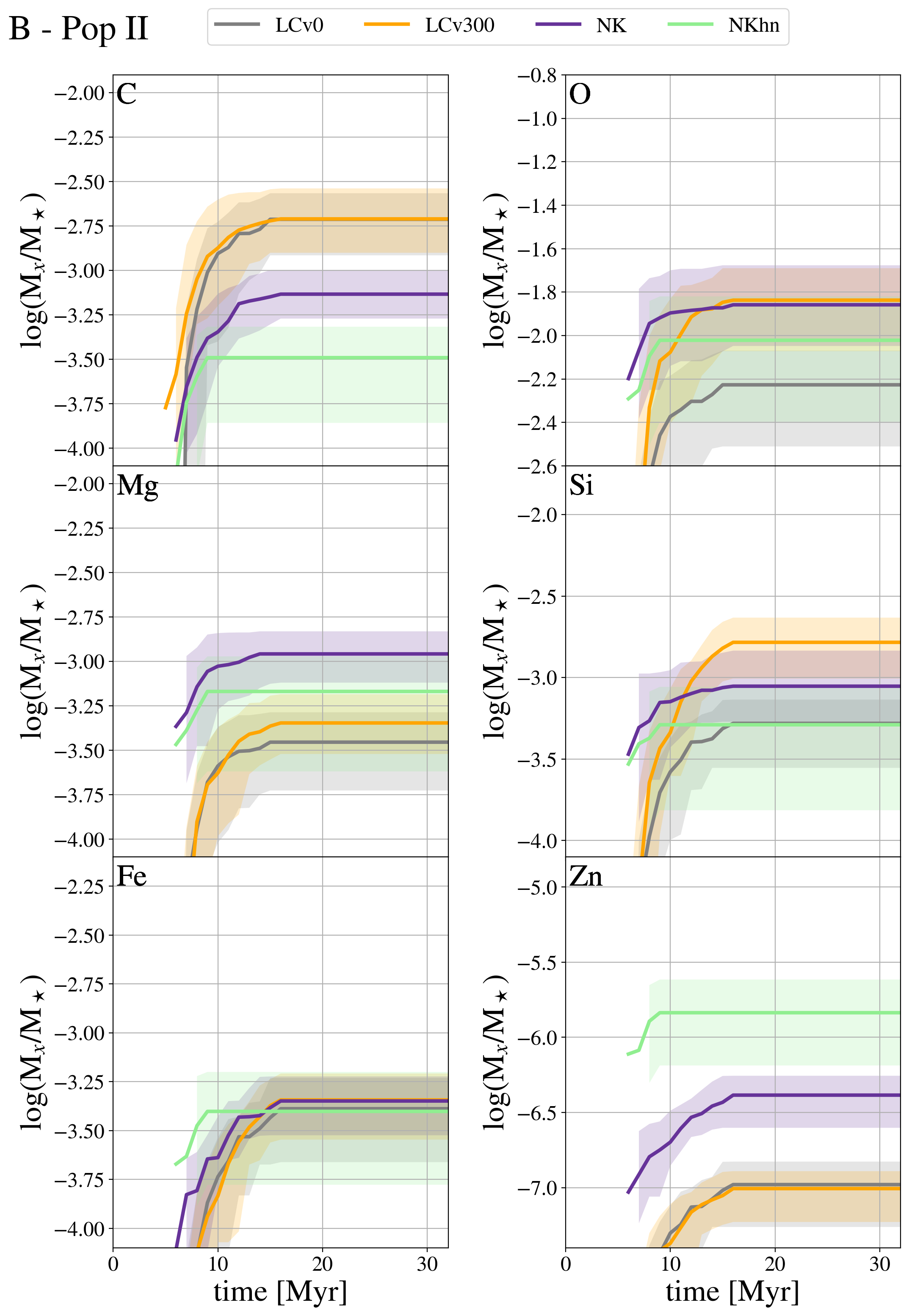}
  \end{subfigure}
       \caption{
     Temporal evolution of the mass return of some elements (from top left to bottom right: carbon, oxygen, magnesium, silicon, iron, and zinc) according to different sets of yields, normalized to the mass of the burst  ($M_\star=10^3\mbox M_\odot$). 
     Solid lines and shaded areas represent the median and the 16/84 percentiles over 50 realizations. 
     \textit{Panel A}: Pop III  model considering SNe yields from \citet{nomoto13} with (green) and without (purple) hyper-novae, and the all-energy model from \citet{heger10} (red).
     \textit{Panel B}: Pop~II/I model considering $v = 0 \,\rm km/s$ (gray) and $v =300 \,\rm km/s$ (yellow) SNe yields from \citet{limongi18}, and yields from \citet{nomoto13} with (green) and without (purple) hyper-novae. \addition{The sets of yields from \citet{heger10} and \citet{limongi18} with $v = 0 \,\rm km/s$ are part of the fiducial model for Pop~III and Pop~II/I, respectively (\Cref{tab:fiducial_parameters}). }
     }
     \label{fig:figure_4}
\end{figure*}

\begin{figure}
    \centering
    \includegraphics[width=0.5\textwidth]{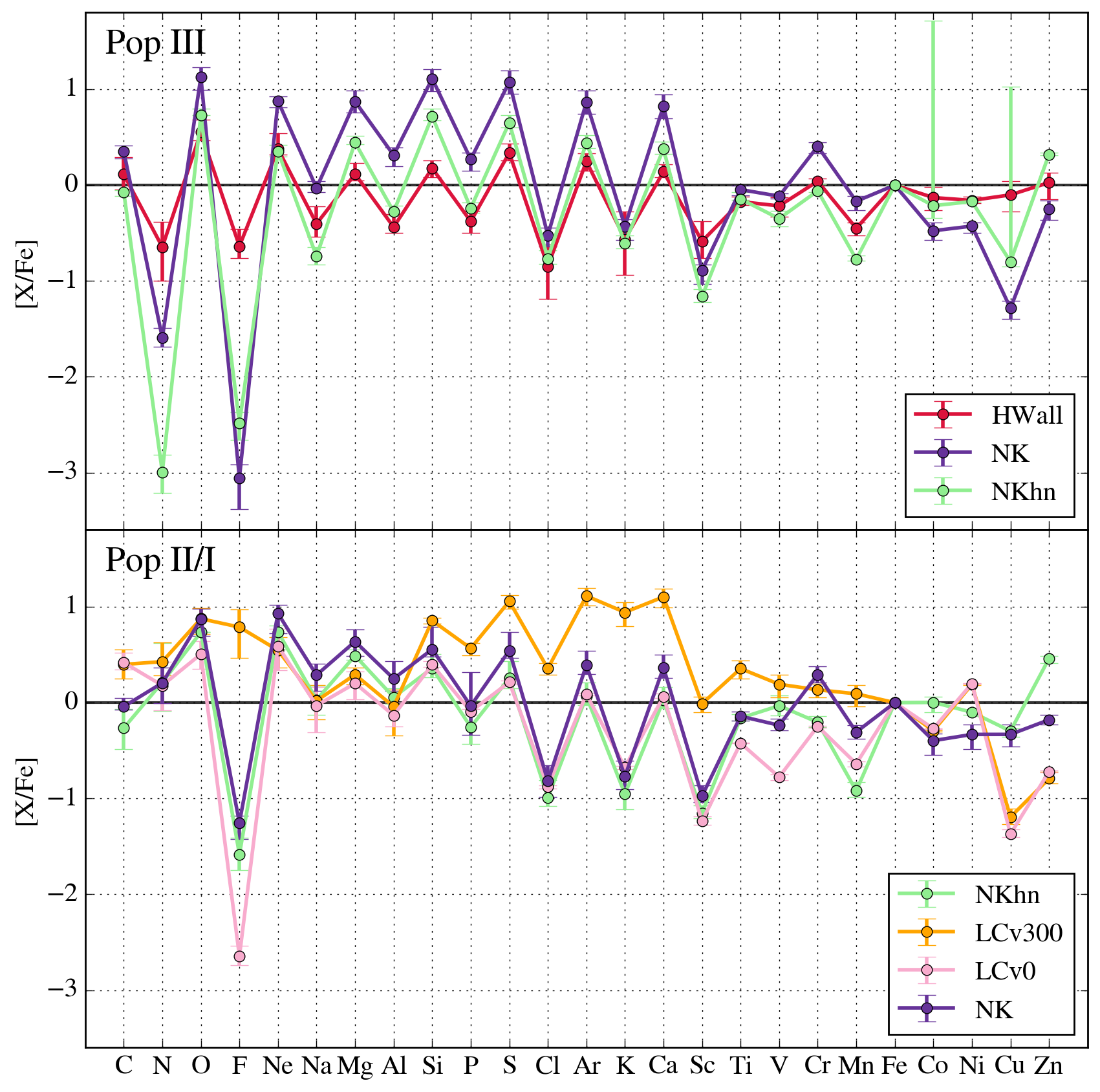}
    \caption{Predicted chemical abundances of elements between carbon and zinc, for a stellar mass burst of $10^3 \rm M_\odot$ after 30 Myr after the initial burst for different sets of yields (see legend). Points and error bars are the median and 16/84 percentiles over 50 realizations \addition{while the two panels refer to the Pop~III (top) and Pop~II/I model (botom)}.}
    \label{fig:figure_5}
\end{figure}

The predictions of chemical evolution models are strongly influenced by the choice of stellar yields, which remain affected by substantial theoretical uncertainties \citep{romano10,cote17}. Here, we explore the impact of different sets of stellar yields on the resulting ISM chemical enrichment, adopting the same initial conditions and total stellar mass ($M_{\star} = 10^3 M_{\odot}$) for all simulations and for both Pop~III and Pop~II/I star formation. In all cases, the stellar populations are generated through the random sampling of the IMF, as described in Sec. \ref{sect:stochasticIMF}.

\Cref{fig:figure_4} shows the time evolution of the total mass of six representative elements (C, O, Mg, Si, Fe, and Zn) in the ISM, normalized to the stellar burst mass. Panel~A focuses on the Pop~III case, comparing yields from \cite{heger10} including all SN energies (HWall), and from \cite{nomoto13} with (NKhn) and without (NK) hypernovae contribution. We refer the reader to \Cref{app:HW} for the comparison between different explosion energies. For C, O, and Fe, the predicted evolution is broadly consistent across all models, with differences confined within $\lesssim0.5$~dex over the first 30 Myr. This indicates that the different stellar models yield broadly consistent abundance ratios for these elements under the adopted assumptions.
In contrast, Mg and Si show stronger model dependencies. Overall, the NK yields predict systematically higher ejected elemental masses compared to HWall, with discrepancies up to $\sim 0.7$ dex. These offsets arise from the different assumptions adopted in the two yield sets, particularly regarding the internal structure of the progenitor stars and the location of the mass cut in the explosion models. The largest discrepancies occur for Zn, whose production is highly sensitive to the adopted SN physics. The NKhn model produces the highest Zn masses, followed by HWall, while the standard NK set underproduces zinc by nearly 1~dex relative to NKhn at $t \approx 30$~Myr. These differences highlight the significant impact of hypernovae and explosion energetics on the yields of iron-peak elements in metal-free stars.

Panel~B of Fig.\ref{fig:figure_4} shows analogous results for Pop~II/I stars. Elements such as Fe exhibit remarkable robustness: the predicted $\log(M_{\rm Fe}/M_\star)$ varies by less than 0.2 dex across all models, suggesting that Fe production is relatively robust across the set of Pop~II/I yield prescriptions explored. In contrast, C is strongly model-dependent: models from \citet{limongi18} (LCv0 and LCv300) predict significantly higher carbon production ($\log(M_{\rm C}/M_\star) \approx -2.75$) compared to the \citet{nomoto13} models ($\approx -3.50$), due to the enhanced internal mixing and mass loss along with the adoption of different $^{12}$C($\alpha$, $\gamma$)$^{16}$O rates. 
Mg and Si show a moderate dependence on the assumed stellar rotation, with LCv300 yields exceeding LCv0 by up to $\sim 0.25$~dex and $\sim 0.5$~dex, respectively. This indicates that rotation can influence the synthesis of $\alpha$-elements, with a non-negligible but element-dependent impact. Indeed, rotation mainly affects the yields of elements that are produced mostly during the hydrostatic burning phases. 
Finally, the Pop~III models show the strongest discrepancies for Zn. The NKhn (NK) model predicts $\gtrsim1$ ($0.5$) dex higher Zn abundance compared to LCv300 and LCv0. This confirms that Pop~II/I Zn production is highly sensitive to both the explosion energy and the stellar yields models.

To examine the global chemical signature, \Cref{fig:figure_5} shows the [X/Fe] abundance patterns predicted by different stellar yield sets for elements from C to Zn.
In the Pop~III case (upper panel) a small group of species (N, F, Al, Si, and Cu) exhibits offsets larger than 1 dex between yield sets, whereas elements such as C, Cl, Ne, Ar, Ca, Sc, Ti, V, Mn, and Co remain relatively stable, with differences confined within $\lesssim0.5$ dex. Up to Mn, the NK yields generally predict higher [X/Fe] than both HWall and NKhn, while for heavier elements NK becomes the lowest of the three. This behaviour highlights which elements are most sensitive to the adopted Pop~III yield prescriptions and which are comparatively robust.

For Pop~II/I stars in Fig.\ref{fig:figure_5} (lower panel), the LCv300 yields are those that deviate the most from the other prescriptions. Several elements, most notably F, Cl, Ar, K, Ca, Sc, and Ti, show differences exceeding 2 dex when LCv300 is compared to the remaining models. For Zn, the NKhn yields predict the highest [Zn/Fe] values, followed by NK, while both LCv0 and LCv300 converge to the same abundance regardless of stellar rotation. Carbon displays the opposite behaviour: NK and NKhn predict lower [C/Fe] values than LCv0 and LCv300 by up to $\sim0.5$ dex. Differences of order $\sim 1$ dex also arise for Mn, where LCv0 and LCv300 diverge from each other, as do NK and NKhn. Finally, Cu exhibits one of the largest offsets between the sets of yields, with NK-based yields (NK and NKhn) exceeding the LC models (LCv0 and LCv300) by more than $\sim 1$ dex. Although some elements, such as O and Ne, remain comparatively less sensitive to the adopted model, these trends show that stellar rotation (LCv300) and hypernova contributions (NKhn) are the primary drivers of scatter in Pop~II/I yield predictions.

%---------------------- PopIII- pre-enrichment ----------------------
\subsection{Pop~III pre-enrichment}\label{sec:preenrichment}
\begin{figure*}
    
    \centering
    \includegraphics[width=1\textwidth]{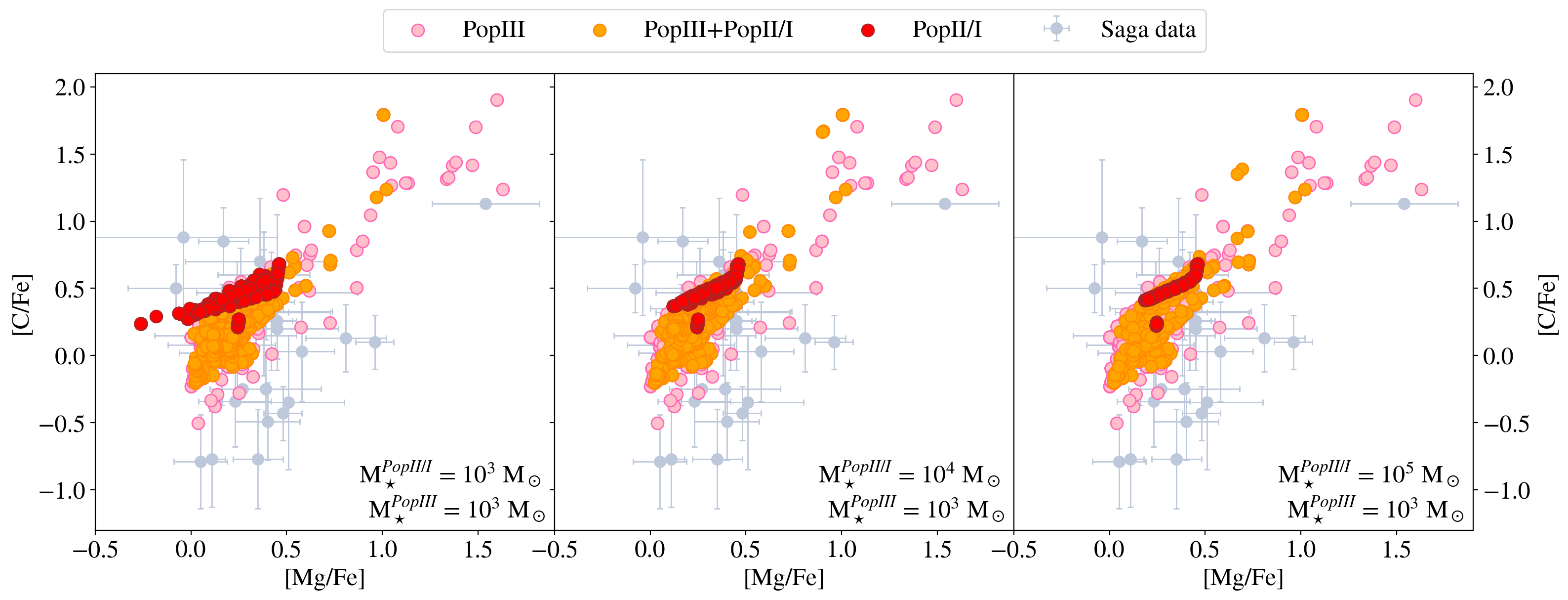}
    \caption{[C/Fe]-[Mg/Fe] diagnostic maps in the case of a single Pop~III (pink) or Pop~II/I (red) stellar burst, and for a combination of a Pop~III and a Pop~II/I burst (PopII+PopIII, orange). Colored points represents the gas composition at a given time (here within the first 30 Myr) \addition{with a time resolution of 1 Myr over a total of 50 realizations}. The burst mass of Pop~III is fixed at $10^3 \rm M_\odot$ in each panel, while for Pop~II/I it varies between panels, as indicated in the bottom right corner. Gray points are the same as in \Cref{fig:figure_2}.}
    \label{fig:preenrichment-diagnostic}
\end{figure*}

\begin{figure}
    \centering
    
    \includegraphics[width=0.49\textwidth]{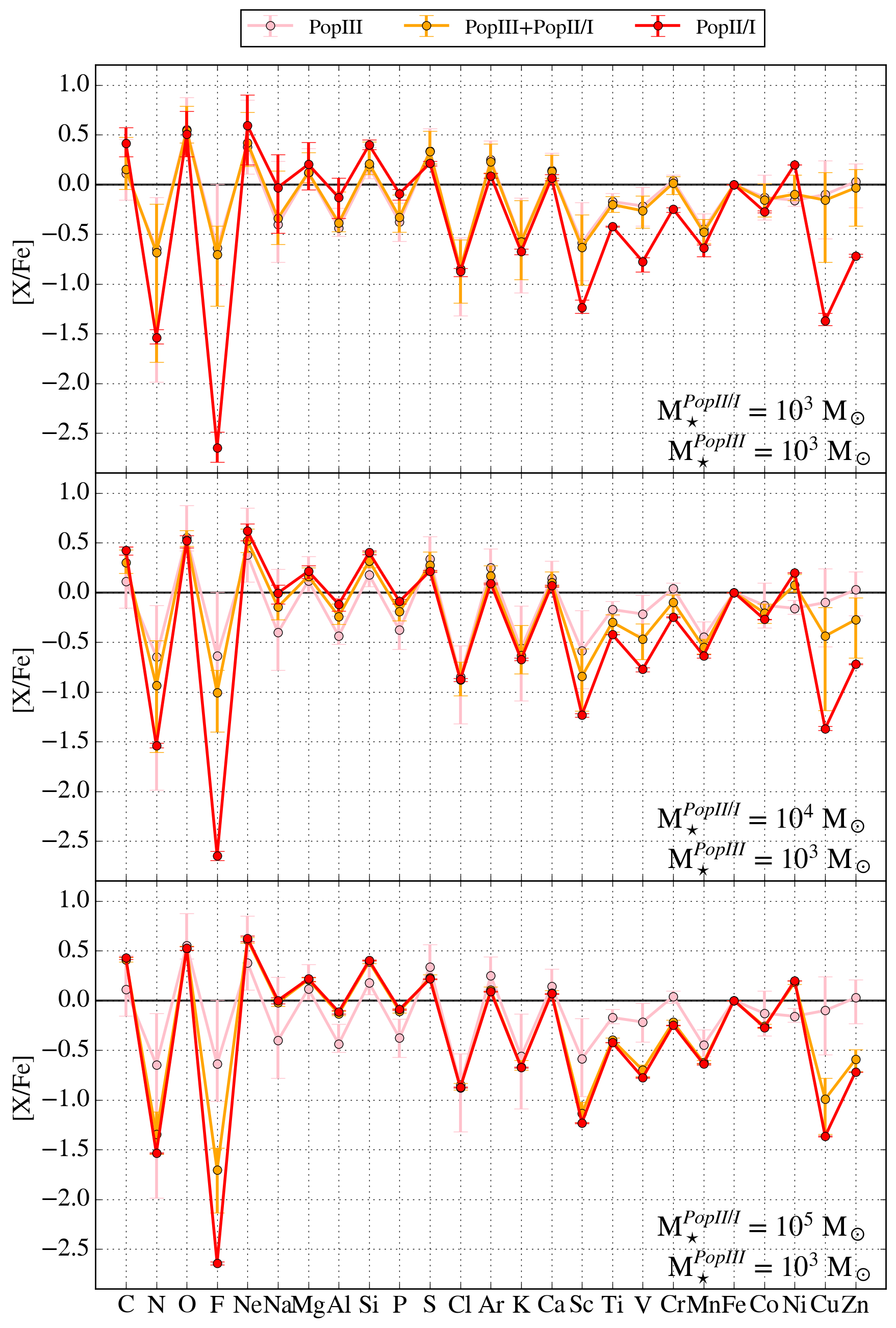}
    \caption{Predicted chemical abundance patterns 30 Myr after the initial burst. Models are the same as in \Cref{fig:preenrichment-diagnostic}. 
    \addition{Points and error bars represent the median and 2.5/97.5 percentiles over 50 realizations}.}
    \label{fig:preenrichment-abundance}
\end{figure}
We now examine the impact of Pop~III pre-enrichment, evaluating how the chemical imprint left by an earlier Pop~III episode alters the abundance patterns produced by a subsequent Pop~II/I stellar population. In particular, we assess whether a genuine Pop~III+Pop~II/I enrichment sequence could produce detectable variations in the chemical composition of the gas, distinct from those produced by a single Pop~II/I burst.
To this end, we model the chemical evolution as the combination of two successive bursts, an initial Pop~III star formation episode followed by a Pop~II/I burst occurring 4~Myr later. This delay approximately corresponds to the lifetime of the most massive Pop~III SNe progenitors, ensuring that their ejecta enrich the gas before the onset of Pop~II/I formation. It also represents the timescale on which the ISM metallicity exceeds the critical threshold ($Z_{\rm ISM} > Z_{\rm crit}$) required to trigger Pop~II/I star formation.
The  Pop~III stellar burst mass is fixed at $M_{\star,\mathrm{III}} = 10^{3}\,M_{\odot}$, while the Pop~II/I burst mass, $M_{\star,\mathrm{II}}$, is varied between $10^{3}$, $10^{4}$, and $10^{5}\,M_{\odot}$. This setup is designed to investigate how the relative contribution of PopII/I enrichment affects the chemical signature associated with the preceding PopIII stellar population in the resulting abundance patterns.

In \Cref{fig:preenrichment-diagnostic} we show the gas-phase [C/Fe] vs [Mg/Fe] enrichment predicted by our model for a single Pop~III or Pop~II/I stellar burst, along with the combined Pop~II+Pop~III case. Similarly to \Cref{fig:figure_2}, each point corresponds to the composition of the ISM between the formation of the stellar burst, $t_0$, and $t_0 + 30$ Myrs, with a 1 Myr resolution. Each panel shows the results for a different value of $M_{\star,\mathrm{II}}$.
Across all panels, a clear trend emerges: Pop~III pre-enrichment systematically produces a larger scatter in the [C/Fe]–[Mg/Fe] plane, independent of the Pop~II/I burst mass. Although the pure Pop~II/I tracks naturally tighten as $M_{\star,\mathrm{II}}$ increases (as discussed in Sec. \ref{sec:chemical_evolution}), the pre-enriched Pop~II+Pop~III models preserve a substantially broader distribution. Even when the Pop~II/I burst is two orders of magnitude more massive than the Pop~III one, the initial Pop~III episode still imprints a noticeable dispersion in the abundance ratios. This enhanced spread originates from the stochastic nature of Pop~III enrichment: each realization samples a different subset of Pop~III SNe progenitors, generating different initial chemical conditions before the onset of Pop~II/I star formation. 

\Cref{fig:preenrichment-abundance} displays the full elemental abundance patterns 30 Myr after the Pop~II burst. As $M_{\star,\mathrm{II}}$ increases, the predicted trends progressively collapse onto the Pop~II-only sequence, indicating that the average chemical pattern becomes increasingly dominated by Pop~II/I enrichment.
Importantly, the convergence of the mean abundance pattern toward the Pop~II-only sequence does not imply the disappearance of the Pop~III signature. Even in the most massive Pop~II/I burst considered, the pre-enriched models retain a significantly larger scatter than the corresponding pure Pop~II/I case. This suggests that stochastic dispersion may represent the most robust surviving signature of Pop~III pre-enrichment once subsequent generations of star formation dominate the overall metal budget.

For lower Pop~II/I stellar burst, however, the situation is different: Pop~III pre-enrichment produces clear and element-dependent deviations, reaching $\gtrsim 1 $dex for several species. In this regime, the chemical fingerprints of Pop~III stars remain detectable, even though the Pop~III stellar mass is smaller than that of the following Pop~II/I generation.

These results highlight a key point: in environments with limited star-formation activity, the memory of Pop~III enrichment is long-lived and can strongly distort the emerging chemical pattern. This makes Pop~III pre-enrichment an essential ingredient in the chemical evolution of the faintest galaxies, such as ultra-faint dwarfs, where stochasticity enhances and preserves the signatures of the first stars.

%===================================================================
%                       DISCUSSION AND CONCLUSIONS
%===================================================================
\section{Discussion and Conclusions} \label{sec:conclusions}
\addition{The present implementation of \CRIMSONS is intentionally based on a simplified closed-box description of chemical enrichment, in which gas inflows, outflows, and spatial mixing are neglected. While this assumption limits the range of astrophysical environments that can be modeled self-consistently, it also represents a key advantage of the present analysis. By removing the additional complexity and degeneracies introduced by gas flows and mixing, \CRIMSONS provides a controlled framework in which the individual effects of: (\textit{i}) stochastic IMF sampling, (\textit{ii}) stellar yield prescriptions, and (\textit{iii}) Pop III pre-enrichment can be isolated and quantified. The results presented below should therefore be interpreted primarily as a characterization of these enrichment processes and of their relative impact on chemical abundance patterns, rather than as predictions for the complete chemical evolution of a specific galaxy.}
\addition{Within this controlled setup, we exploited \CRIMSONS to explore three main questions: how IMF sampling affects stellar mass distributions and abundance ratios, how the choice of stellar yields influences enrichment outcomes, and how Pop III pre-enrichment alters the chemical fingerprints of subsequent Pop II/I bursts. Our main findings are summarized below:}
\begin{comment}
In this work we presented \CRIMSONS, a flexible and computationally efficient chemical-evolution framework designed to incorporate three essential ingredients that are often neglected in standard models:
(\textit{i}) stochastic sampling of the IMF,
(\textit{ii}) the flexibility of adopting different stellar yield sets for both Pop~III and Pop~II/I stars, and (\textit{iii}) the pre-enrichment from primordial Pop~III stars.
In this paper we exploited \CRIMSONS to explore three main questions: how IMF sampling affects stellar mass distributions and abundance ratios, how the choice of stellar yields influences enrichment outcomes, and how Pop~III pre--enrichment alters the chemical fingerprints of subsequent Pop~II/I bursts. Our main findings are summarized below
\end{comment}

\begin{itemize}
    \item The assumption of a fully sampled IMF is valid only above a critical total stellar mass, whose value depends on the IMF functional form (\Cref{fig:figure1}).  
    For a Larson IMF with a characteristic mass $m_{\mathrm{ch}}=0.35\,M_\odot$, convergence is reached for $M_\star\sim10^{5}\,M_\odot$, whereas in top-heavy cases  ($m_{\mathrm{ch}}\sim\,10 M_\odot$), the same level of completeness requires $M_\star \gtrsim 10^6\,M_\odot$.
    For lower stellar masses, the discreteness of the sampling process produces significant deviations in the \emph{effective} IMF and therefore in the resulting chemical enrichment.
    \item For low $M_\star$, IMF sampling naturally introduces scatter in elemental ratios, driven by the random presence or absence of massive SNe progenitors. In the case of top-heavy IMFs, such as those usually assumed for Pop III stars, the scatter in the abundance ratios can be larger than 1 dex, compared to Pop II-like bottom-heavy IMFs (\Cref{fig:figure_2} and \Cref{fig:figure_3}).
    \item The choice of stellar yields leads to systematic and element-dependent differences in predicted chemical abundance patterns (\Cref{fig:figure_4} and \Cref{fig:figure_5}).
    \begin{comment}
        For Pop~III stars, the yields from \citet{heger10} and \citet{nomoto13} produce broadly consistent trends for C, O, and Fe, but diverge for $\alpha$-- and iron--peak elements such as Mg, Si, and Zn with differences reaching up to $\sim 1$\,dex (see panel A of \Cref{fig:figure_4}).
    \end{comment}
    \addition{For Pop~III stars, differences between the yields from \citet{heger10} and \citet{nomoto13} remain within $\lesssim0.5$ dex for C, O, and Fe, but reach up to $\sim1$ dex for elements such as Mg, Si, and Zn (see panel A of \Cref{fig:figure_4})}.
    For Pop~II/I stars, Fe production is remarkably robust (variations $<0.2$\,dex across models), whereas light and iron--peak elements remain strongly affected by rotation and the inclusion of hypernovae (\Cref{fig:figure_4}, panel B). 

    \item The chemical pre-enrichment of a single Pop~III star formation episode can induce a dispersion in chemical abundances (see \Cref{fig:preenrichment-diagnostic}), particularly at early times. \addition{For low-mass Pop~II/I bursts ($M_{\star,\mathrm{II}}\lesssim10^4,M_\odot$), deviations of $\gtrsim1$ dex persist across several elements, whereas at larger Pop~II/I masses the mean abundance pattern progressively approaches the Pop~II/I-only case, while retaining an enhanced realization-to-realization scatter (\Cref{fig:preenrichment-abundance}).}
\end{itemize}
\addition{Our results on stochastic IMF sampling are consistent with previous studies showing that the discrete sampling of the IMF becomes increasingly important in low-mass stellar populations \citep[e.g.,][]{debennessuti17,applebaum21}. In particular, \CRIMSONS recovers the increasing realization-to-realization scatter toward decreasing stellar mass, while explicitly showing how the stellar mass required for convergence depends on the adopted IMF shape. For the configurations explored here, a Pop II/I-like Larson IMF approaches complete sampling at \(M_\star\sim10^{4}-10^{5}\,M_\odot\), whereas the top-heavy Pop III configurations require \(M_\star\gtrsim10^6\,M_\odot\).} \addition{Recent cosmological hydrodynamic simulations of UFD analogs further demonstrate that the treatment of IMF sampling can affect not only chemical enrichment but also the resulting star-formation histories and stellar metallicities \citep[e.g.,][]{jeon26}. In particular, \cite{jeon26} find significant differences in the stellar masses and metallicities of their simulated UFDs when adopting burst, stochastic, or individual-star IMF sampling schemes. Their results therefore complement the controlled experiments performed with \CRIMSONS, illustrating the astrophysical consequences that different treatments of IMF discreteness can have once stellar feedback and galaxy evolution are modeled self-consistently.}

\addition{The sensitivity of the predicted chemical patterns to the adopted stellar yields is also qualitatively consistent with previous chemical-evolution studies, which have shown that uncertainties in stellar nucleosynthesis prescriptions propagate into element-dependent differences in predicted abundances \citep[e.g.,][]{romano10,romano14,romano20}. A direct quantitative comparison is not straightforward because these studies couple stellar yields to a complete chemical-evolution model, including a star-formation history and gas evolution, whereas \CRIMSONS isolates the ejecta produced by individual stellar populations within the simplified closed-box setup. Nevertheless, the qualitative conclusion is the same: some elements are comparatively robust against the adopted yield prescription, whereas others retain a strong dependence on stellar-evolution and explosion assumptions.}
% implications
\begin{comment}
   The results presented here have several implications for the study of early galaxy evolution.  
First, our work confirms that the large chemical scatter observed among the most metal--poor systems may largely reflect stochastic sampling of the IMF rather than complex assembly histories.  
Second, uncertainties in yield prescriptions and IMF sampling can introduce systematic abundance variations of $\sim 0.5$--1\,dex, comparable to or larger than observational errors, and should therefore be propagated in chemical evolution modeling.  
Third, the persistence of Pop~III signatures in low-mass systems suggests that present-day UFDs remain promising sites to constrain the nature of the first stars, provided that models explicitly accounts for discreteness effects. 
\end{comment}

\addition{The results presented here have important implications for the interpretation of abundance patterns in metal-poor stars and low-mass systems. In particular, a large abundance scatter can arise naturally from the discrete sampling of a single underlying IMF when only a limited stellar mass is formed, without requiring intrinsic variations in the IMF itself. In the poorly sampled Pop~III case explored here, the resulting dispersion exceeds 2 dex in [C/Fe] and 1.5 dex in [Mg/Fe]. Moreover, the predicted abundance patterns reflect the combined effects of IMF sampling and the adopted stellar yields, implying that individual abundance ratios should not be interpreted independently of these modelling uncertainties.
Pop~III pre-enrichment introduces an additional complexity in this interpretation. At low Pop~II/I stellar masses ($M_{\star,\mathrm{II}}\lesssim10^4,M_\odot$), the initial Pop~III enrichment can produce deviations exceeding 1 dex for several elements, while at larger Pop~II/I masses its most persistent signature is an enhanced realization-to-realization scatter. Consequently, the interpretation of chemically peculiar metal-poor stars and low-mass systems as tracers of Pop~III enrichment should account for both the stochasticity of the enriching stellar population and the subsequent contribution from Pop~II/I stars.}

% limitations
\addition{The current version of \CRIMSONS\ neglects inflows, outflows, and spatial mixing, which are expected to reduce the amplitude of abundance scatter and to modulate the persistence of Pop~III signatures. }
The model also assumes fixed delay times and does not explicitly treat binaries or rotation-dependent lifetimes.  
Future extensions will include multiple bursts and gas flows, allowing a direct comparison with cosmological simulations and observed dwarf galaxies.  
In addition, expanding the framework to include updated yield grids will improve predictions for the most uncertain elements (e.g. Zn, Ti, Al) along with the implementation of n-capture elements (Eu, Ba, La, Y, Ce). 

From an observational standpoint, the results motivate (i) high-resolution abundance studies in the faintest galaxies to quantify the element-to-element scatter, (ii) statistical analyses testing consistency with stochastic IMF sampling, and (iii) joint modeling of abundance ratios and stellar masses to infer typical star-formation burst sizes in UFDs.  
Such comparisons will enable stringent tests of early enrichment models and the nature of the first stars.

% conclusion
In summary, our study highlights that stochastic IMF sampling and Pop~III pre-enrichment are essential components of early chemical evolution.  
These processes, together with yield prescriptions, introduce abundance variations comparable to observational uncertainties, and can dominate the chemical enrichment in low-mass star-forming systems. 
Therefore, realistic modeling of the earliest chemical signatures, and their connection to observed metal-poor stars, requires explicit treatment of IMF discreteness and the inclusion of Pop~III contributions.  
We release \CRIMSONS\ \addition{both as a publicly available web-based tool and a Python package,} to facilitate such investigations and to help quantify uncertainties in future chemical-evolution studies.

\begin{acknowledgements}
      This project has received funding from the European Research Council under the European Union’s Horizon 2020 research and innovation program (grant agreement No 804240). 
      This project has been funded by the Italian National Institute for Astrophysics (INAF) through the “Progetti di Ricerca Fondamentale 2024” program, under the Mini-Grant GEMS (Grant No. 1.05.24.07.02).
      S.S. and I.V. acknowledge support from the PRIN-MIUR2017, prot. n. 2017T4ARJ5. 
      D.R. and M.R. acknowledge support by INAF through the program “Finanziamento della Ricerca Fondamentale”, project \emph{“An in-depth theoretical study of CNO element evolution in galaxies”} Fu.~Ob.~1.05.12.06.08 (PI: D.~Romano).
\end{acknowledgements}

%---------------------- BIBLIOGRAPHY ----------------------
\bibliographystyle{aa}
\bibliography{mybib}

%---------------------- APPENDIX ----------------------
\begin{appendix}
\appendix

%===================================================================
%                       APPENDIX A
%===================================================================

\section{Time evolution of [C/Fe] and [Mg/Fe] } \label{app:time_sampling}

\begin{figure}
    \centering
    \includegraphics[width=1\linewidth]{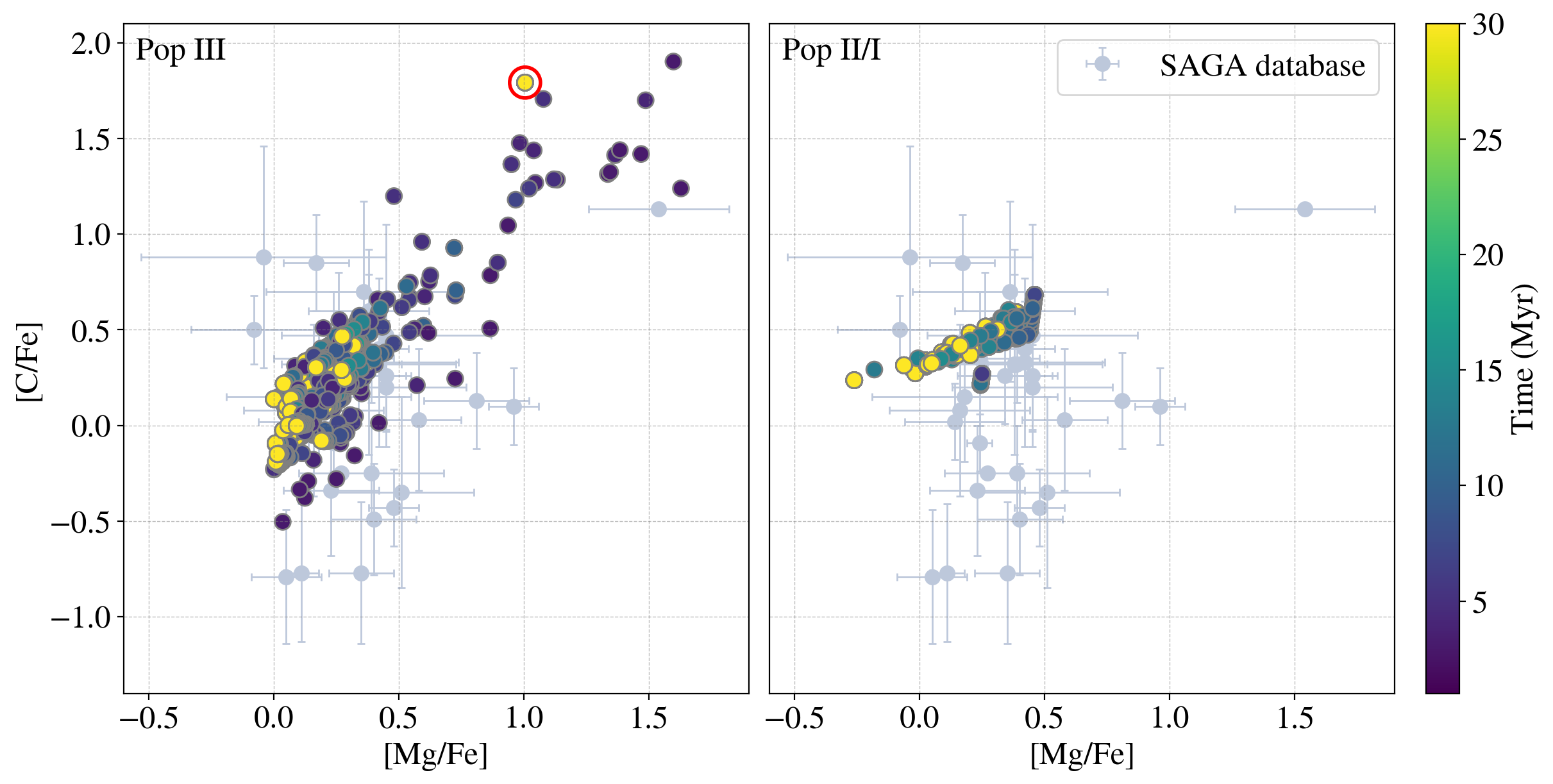}
    \caption{ \addition{Diagnostic map of [C/Fe] versus [Mg/Fe] for Pop~III (left) and Pop~II/I (right) enrichment predicted by our fiducial model, similar to \Cref{fig:figure_2}. Points correspond to the gas-phase abundance at different times within the first 30 Myr after the stellar burst for each of the $50$ realizations and are colored by the time after the stellar burst. The circled point indicates a realization in which the sampling produces only one star within $10 \msun$ and $ 100 \msun$, resulting in a single enrichment event.}}
    \label{fig:time_sampling}
\end{figure}

The time evolution of the chemical abundances depends on the mass of the stars enriching the gas. \Cref{fig:time_sampling} shows the evolution of [C/Fe] versus [Mg/Fe] for a poorly-sampled IMF and for the first 30 Myrs from the stellar burst. At early times (<5 Myr) the abundances are dominated by enrichments of rare massive stars, producing high abundance ratios, specially in the Pop~III case. At later times (> 10 Myr), the majority of SNe explode decreasing the abundance ratios towards the solar value ([X/Fe] = 0). 
One notable exception is the point circled in red in the Pop~III case. In that specific realization, the random sampling produced only one star within $10 \msun$ and $ 100 \msun$. That sole star enriches the gas within the first 30 Myrs, resulting in fixed abundance ratios of [C/Fe] = 1.79 and [Mg/Fe] = 1.02.

%===================================================================
%                       APPENDIX B
%===================================================================
\section{Energy dependent Pop~III stars yields} \label{app:HW}

\begin{figure}
    \centering
    \includegraphics[width=1\linewidth]{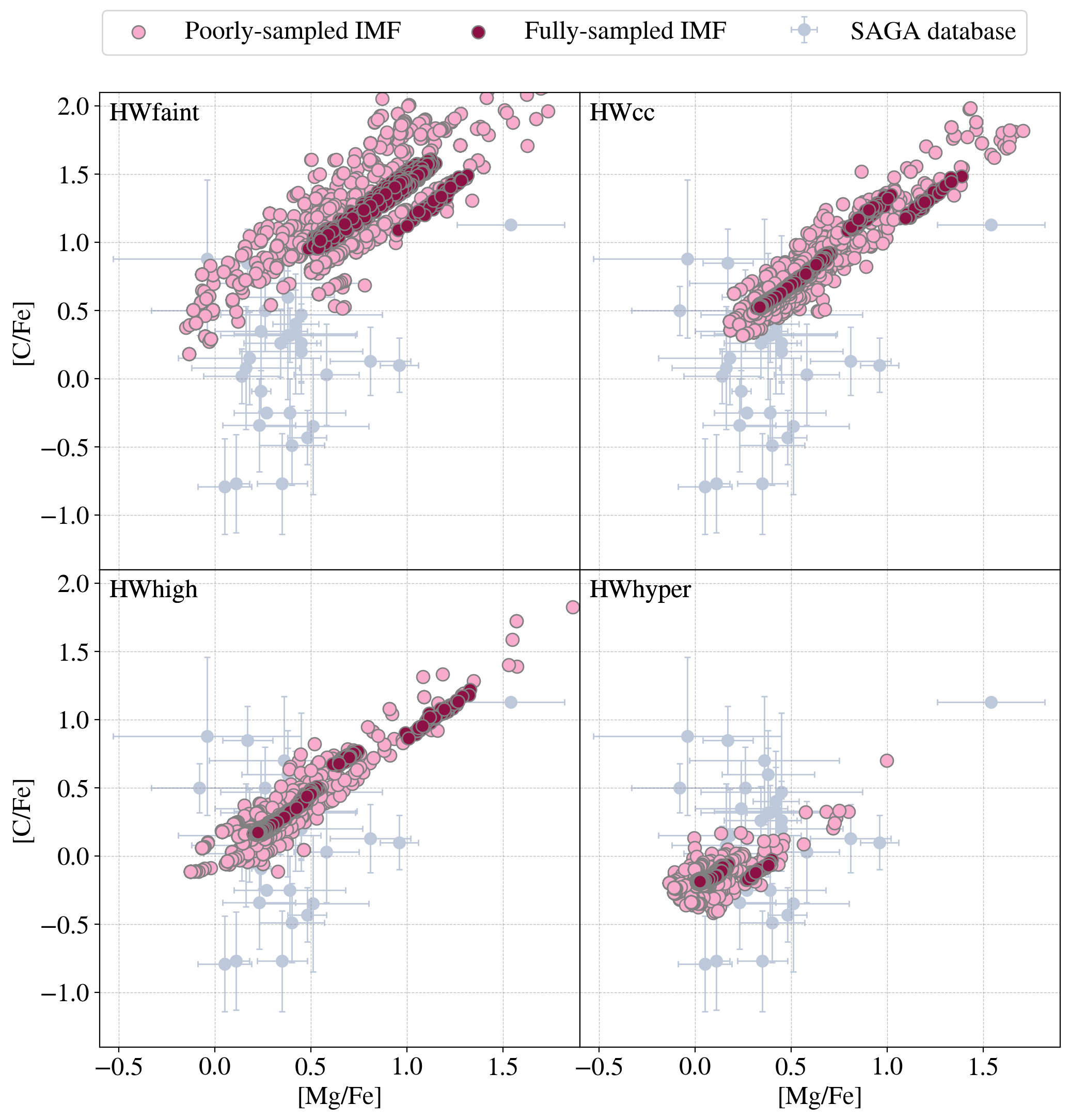}
    \caption{Identical to \Cref{fig:figure_2}, with panels showing different SN energies, as indicated in the top left of each panel. }
    \label{fig:sampling_energy_dependency}
\end{figure}

\begin{figure}
    \centering
    \includegraphics[width=\linewidth]{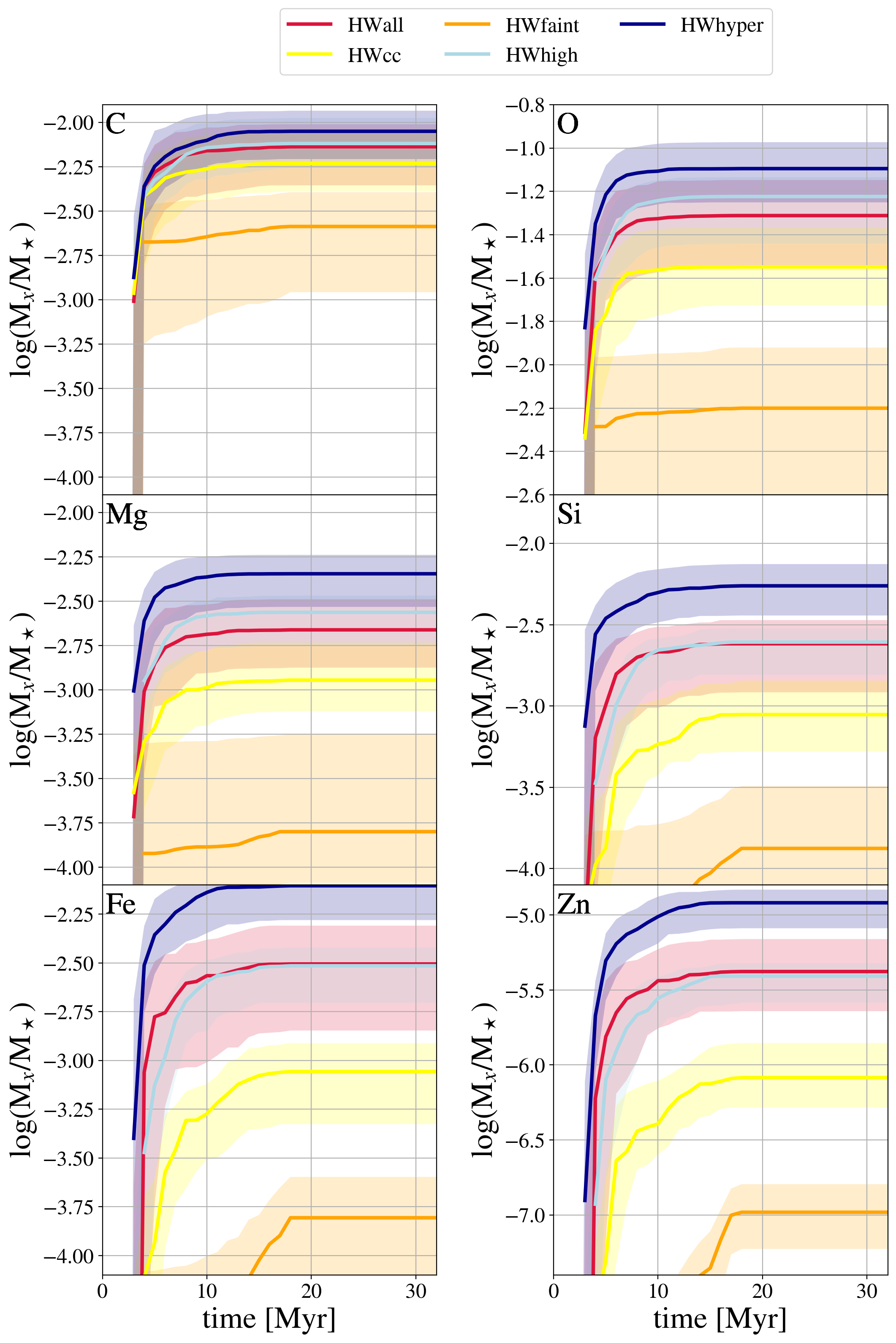}
    \caption{The evolution of the total masses of six elements (panels), normalized to the stellar burst mass, for different Pop~III SN explosion energies. Solid lines show the median values from 50 realizations, while the shaded regions indicate the 16th--84th percentile range. Colors correspond to different explosion-energy models: \textit{HWfaint}, \textit{HWcc}, \textit{HWhigh}, and \textit{HWhyper}. The model labeled \textit{HWall} assumes a uniform 25\% probability for each explosion-energy channel and \addition{represents our fiducial model}.}
    \label{fig:Mx_HW}
\end{figure}

\begin{figure}
    \centering
    \includegraphics[width=\linewidth]{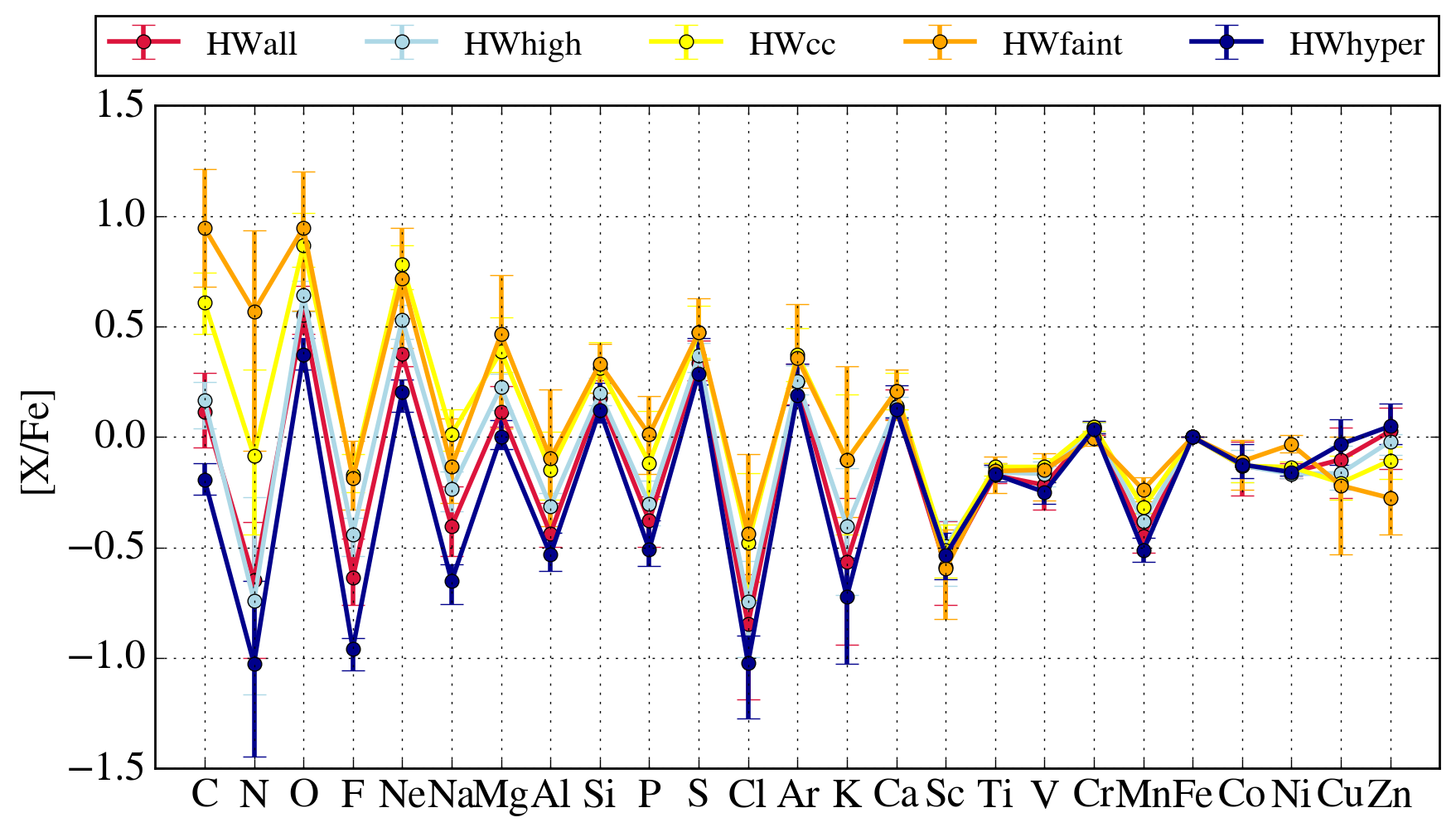}
    \caption{Predicted chemical abundances, [X/Fe], of elements from C to Zn at 50~Myr after a Pop~III stellar burst with $M_\star = 10^3 \msun$. Colors follow the convention as \Cref{fig:Mx_HW}.}
    \label{fig:average_HW}
\end{figure}

The explosion energies of Pop~III supernovae in the progenitor mass range $m_\star = 10$--$100\,\msun$ remain highly uncertain. In this section, we investigate how different SN explosion energies affect the chemical enrichment produced by Pop~III stellar populations. To this extend, we explore the explosion energies as defined in \citet{heger10}: : \textit{-faint}, with explosion energies $\rm E_{SN} = [0.3$--$0.6]\times10^{51} \msun \erg$; \textit{-cc} (core-collapse), with $\rm E_{SN} = [1.2$--$1.5]\times10^{51} \erg$; \textit{-high} (high-energy SNe), with $\rm E_{SN} = [1.8$--$3.0]\times10^{51} \erg$; and \textit{-hyper} (hypernovae), with $\rm E_{SN} = [5.0$--$10.0]\times10^{51} \erg$. The \textit{-all} model assumes an equal probability (25\%) for each explosion-energy channel.

\Cref{fig:sampling_energy_dependency} is analogue to \Cref{fig:figure_2} and each panel corresponds to a different explosion energy for the Pop~III SNe. The impact of the sampling of the IMF is evident in all four cases, with a scatter between realizations that spans 2 dex  in the faint SNe case and 1 dex in the hypernovae case. The presence of a scatter regardless of the explosion energy confirms that it is not driven by the mixed explosion energies assumed in the fiducial model but it is a effect of the incomplete sampling of the IMF. 

\Cref{fig:Mx_HW} shows the time evolution of the total mass of six representative elements, normalized to the stellar burst mass, for different Pop~III explosion-energy models. Higher explosion energies systematically lead to larger overall metal enrichment. The absolute differences between the models increase toward heavier elements: for example, the enrichment differs by roughly one order of magnitude for O, while reaching nearly two orders of magnitude for Zn between the faint-SN and hypernova models.

Lastly, \Cref{fig:average_HW} presents the predicted [X/Fe] abundance patterns for all elements from C to Zn at 30~Myr after the Pop~III burst. While the total metal enrichment strongly depends on the explosion energy, the abundance ratios are comparatively less sensitive for most intermediate and iron-peak elements. The largest variations are instead found among the light elements, especially C and N, where differences between explosion-energy models can exceed 1~dex.

%===================================================================
%                       APPENDIX C
%===================================================================

\section{Chemical abundance after 15 Myrs} \label{app:enrichment_15}

\begin{figure}
    \centering
    \includegraphics[width=1\linewidth]{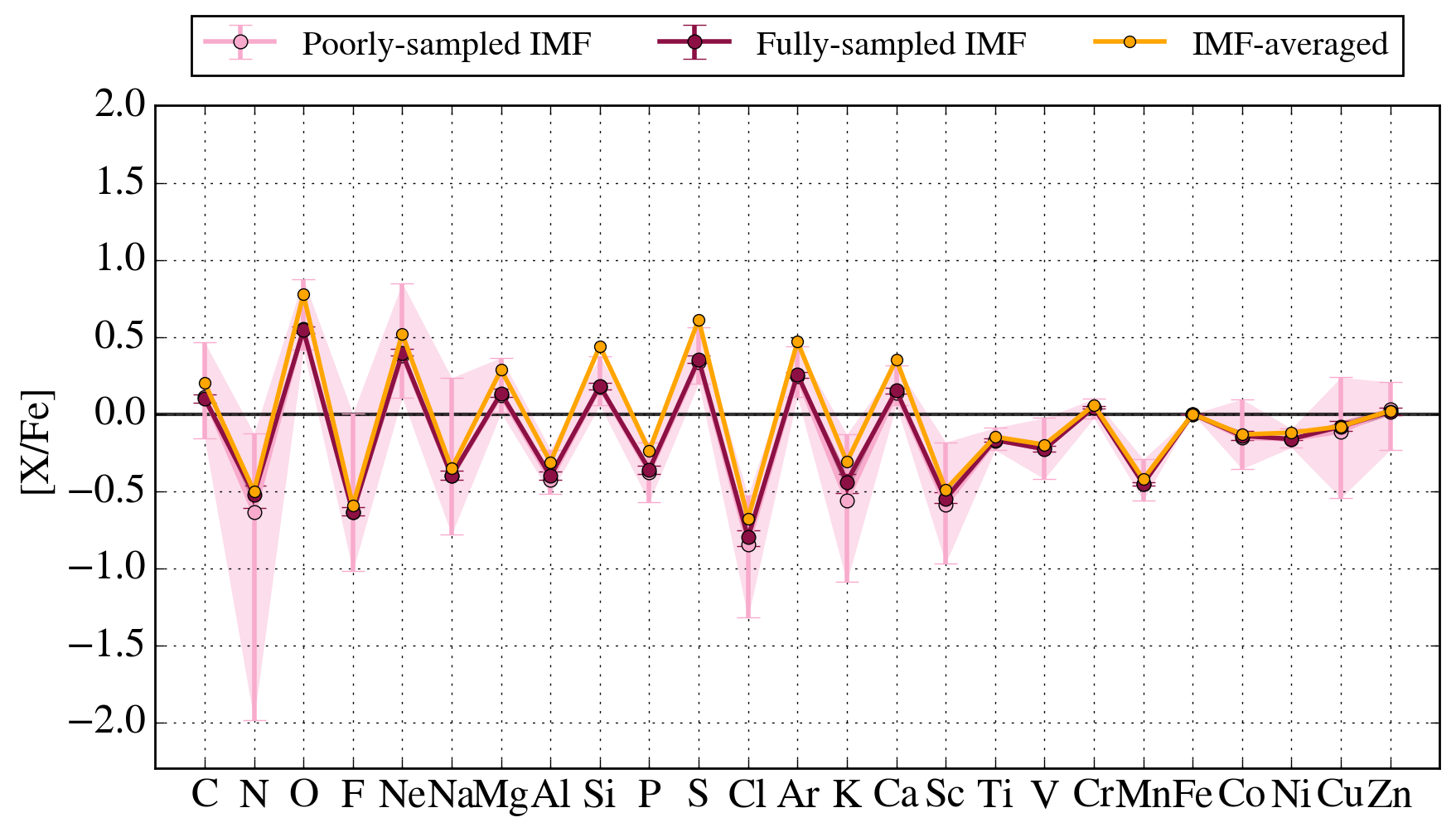}
    \caption{Gas-phase abundance ratios, [X/Fe], for elements from C to Zn measured 15~Myr after a Pop~III stellar burst. The figure compares the chemical enrichment predicted for a poorly sampled IMF, a fully sampled IMF, and the corresponding IMF-averaged yields. \addition{Error bars and shaded} regions indicate the scatter as 16th--84th percentile across the 50 realizations in the sampled cases.}
    \label{fig:abundance_15Myr}
\end{figure}

The chemical enrichment pattern depends on which stars have already exploded as SNe and therefore evolves with time after the initial starburst. \Cref{fig:abundance_15Myr} shows the abundance patterns 15~Myr after a Pop~III star formation for the poorly sampled IMF, fully sampled IMF, and IMF-averaged models. Compared to the enrichment pattern at the Pop~III/Pop~II transition threshold discussed in \Cref{sec:IMF_sampling_chemical_enrichment} and shown in \Cref{fig:figure_3}, the differences between the sampled and IMF-averaged models are significantly reduced at this later stage. By 15~Myr, lower-mass supernova progenitors have started to dominate the enrichment, decreasing the impact of stochastic IMF sampling. Since lower-mass stars are more numerous, their contribution is less sensitive to the discrete nature of the IMF realization, leading to a closer agreement between the sampled and IMF-averaged models.

For the same reason, the realization-to-realization scatter in the sampled models is also reduced relative to the earlier evolutionary stage. Nevertheless, noticeable differences remain for light elements whose production is particularly sensitive to specific stellar mass ranges. In particular, elements such as C and N, still exhibit $\sim$1 dex scatter in the poorly sampled IMF model, highlighting that IMF averaging can continue to bias the predicted abundance patterns in low-mass and poorly star-forming environments, even after the enrichment becomes dominated by lower-mass supernovae.

%===================================================================
%                       APPENDIX D
%===================================================================

\section{Time evolution of [C/Fe] and [Mg/Fe] with Popp~III preenriched} \label{app:time_evolution_preenrichment}

\begin{figure*}
    \centering
    \includegraphics[width=1\linewidth]{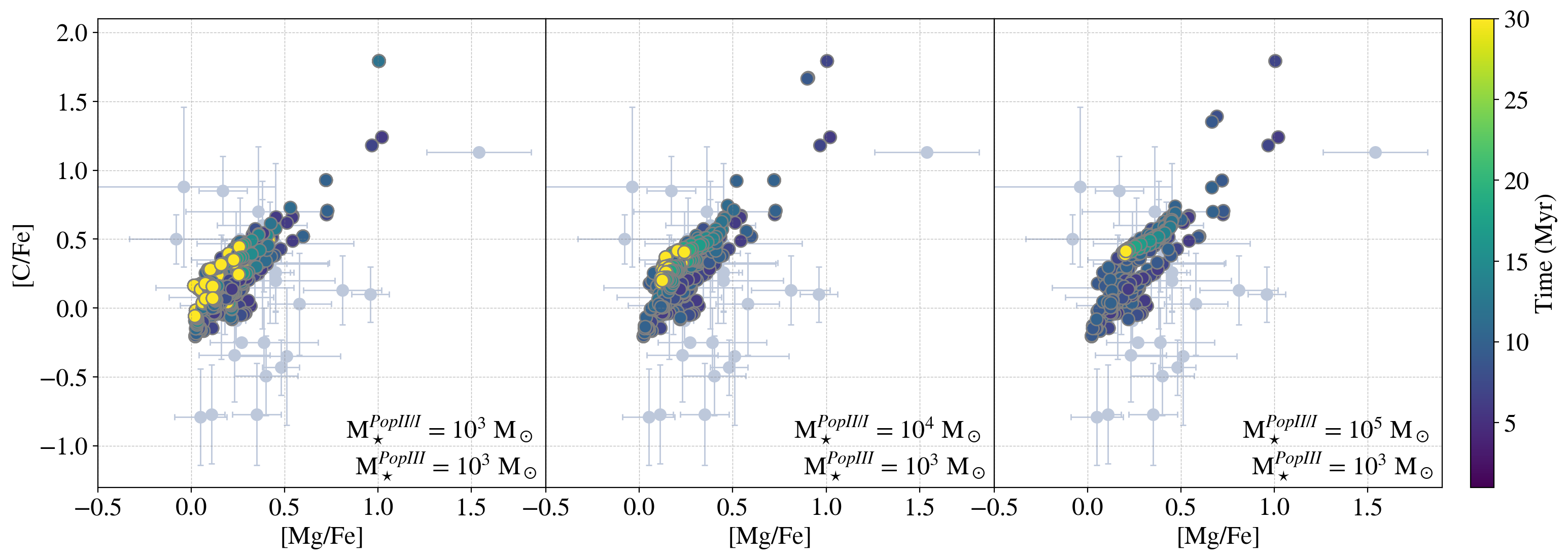}
    \caption{ \addition{Diagnostic map of [C/Fe] and [Mg/Fe] for different combinations of Pop~III and Pop~II/I enrichment, as indicated in the bottom right of each panel. Each panel shows the time evolution of the chemical abundance released in the ISM by $50$ realizations from the formation of the Pop~II/I stars with a time resolution of $1$ Myr. The gray points are take from the SAGA database \citep{saga_database} and corrected for carbon depletion \citep{placco14}.}}
    \label{fig:time_preenrichment}
\end{figure*}

\addition{The ISM chemical abundances evolve with time based on the mass of the stars enriching the gas. As a result, when considering a Pop~III pre-enrichment, the time evolution depends on the mass of the stellar population formed. Figure~\ref{fig:time_preenrichment} shows the time evolution of [C/Fe] and [Mg/Fe] for the three combinations of Pop~III and Pop~II enrichment (panels), illustrating the results for $50$ realizations. The mass of Pop~III stars formed is fixed at $10^3 \msun$, while the Pop~II stellar mass increases left to right from $10^3 \msun$ to $10^5 \msun$. Such combinations coincide with the ones of Sect. \ref{sec:preenrichment}. The temporal evolution is represented by the color scale, tracing the time elapsed since the formation of the Pop II stars.
When the Pop~III and Pop~II/I masses are comparable, [C/Fe] and [Mg/Fe] are within 0 and 0.5 after 30 Myrs, with a large scatter. Increasing the mass of Pop~II/I stars formed reduces the scatter, as evident in the last panel of the figure, where the chemical abundance after 30 Myrs converges for all $50$ realizations at [C/Fe] $= 0.4$ and [Mg/Fe] $= 0.2$. Such values are consistent with the enrichment from Pop~II/I in the case of a fully sampled IMF (see Fig.~\ref{fig:figure_2}) and further support the result that after $30$ Myrs the chemical signatures of Pop~III pre-enrichment are washed away when the Pop~II/I dominates.}  

\end{appendix}
\end{document}

%% file: include_tex/pacchetti.tex
\usepackage[english]{babel}
\usepackage{amsmath}
\usepackage{amssymb}
\usepackage{graphicx}
\usepackage{subfigure}
\usepackage[normalem]{ulem}

\usepackage{verbatim}

\usepackage{color}
\usepackage{multirow}
\usepackage{mathtools}
\usepackage{epstopdf}

\usepackage{url}
\usepackage{xcolor}

%% file: include_tex/journals.tex
\def\apjl{ApJL}
\def\apjs{ApJS}
\def\aap{A\&A}